\documentclass[letterpaper,twocolumn,10pt]{article}
\usepackage{usenix}

\usepackage{tikz}
\usepackage{amsmath}

\usepackage{filecontents}

\usepackage{amsmath}
\usepackage{amsthm}
\usepackage{etoolbox}
\usepackage{amssymb}
\usepackage{mathtools}
\usepackage{bm}
\usepackage{bbm}
\usepackage{booktabs}
\usepackage{xspace}
\usepackage{color,soul}
\usepackage{multirow}
\usepackage{multicol}
\usepackage{caption}
\usepackage{enumitem}
\usepackage{verbatim}
\usepackage{microtype}
\usepackage{graphicx}
\usepackage{booktabs}
\usepackage{hyperref}
\usepackage{nicefrac}
\usepackage[normalem]{ulem}
\usepackage{tikz}
\usetikzlibrary{trees}
\usepackage{float}
\usepackage{subcaption}
\usepackage[ruled,noend,linesnumbered,vlined]{algorithm2e}
\usepackage{xcolor,colortbl}
\usepackage{hhline}
\usepackage{tabularx}
\usepackage{listings}
\usepackage{minted}
\usepackage{xcolor}
\usepackage{makecell}
\usepackage{pifont}

\usepackage{array}
\newcolumntype{P}[1]{>{\raggedright\arraybackslash}p{#1}}

\usepackage{pifont}
\usepackage{lipsum} 
\usepackage[available]{usenixbadges}

\usepackage{tikz}

\newcommand{\mysubsubsection}[1]{{\vspace{0.25em}\noindent{\textbf{#1.\xspace}}}}

\newcommand{\myvspace}[1]{\vspace{#1}}

\newcommand{\system}{{\text{HSPMD}}\xspace}
\newcommand{\DeviceGroup}{{\textit{DG}}\xspace}
\newcommand{\DistributedStates}{{\textit{DS}}\xspace}
\newcommand{\DeviceGroupUnion}{{\textit{DG Union}}\xspace}
\newcommand{\DistributedStatesUnion}{{\textit{DS Union}}\xspace}
\newcommand{\HeteroSize}{{\textit{HSize}}\xspace}
\newcommand{\HeteroDim}{{\textit{HDim}}\xspace}

\newcommand{\specialcell}[2][c]{%
	\begin{tabular}[#1]{@{}c@{}}#2\end{tabular}}

\makeatletter
\newcommand{\nakedthanks}[1]{%
  \protected@xdef\@thanks{%
    \@thanks
    \protect\@naked@thanks{#1}%
  }%
}
\newcommand{\@naked@thanks}[1]{%
  \let\thefootnote\relax
  \footnotetext{#1}%
}
\makeatother

\begin{document}
\date{}

\title{Hetu v2: A General and Scalable Deep Learning System with Hierarchical and Heterogeneous Single Program Multiple Data Annotations}

\author{\rm{Haoyang Li}$^\S$ \hspace{1.2em} Fangcheng Fu$^\dag$ \hspace{1.2em} \rm{Hao Ge}$^\S$ \hspace{1.2em} \rm{Sheng Lin}$^\S$ \\ \rm{Xuanyu Wang}$^\S$  \hspace{1.2em} 
 \rm{Jiawen Niu}$^\S$ \hspace{1.2em} \rm{Yuming Zhou}$^\S$ \hspace{1.2em} \rm{Xupeng Miao}$^\S$ \hspace{1.2em} \rm{Bin Cui}$^{\S\ddag}$ 
 \\ [.6em] 
 $^\S$School of Computer Science \& Beijing Key Laboratory of Software and Hardware Cooperative Artificial\\Intelligence Systems, Peking University~~~$^\dag$School of Artificial Intelligence, Shanghai Jiao Tong University \\
 $^\ddag$Institute of Computational Social Science, Peking University (Qingdao)\\
}

\maketitle


\begin{abstract}
The Single-Program Multiple-Data (SPMD) paradigm provides a unified abstraction to annotate various parallel dimensions in distributed deep learning (DL) training. With SPMD, users can write training programs from the viewpoint of a single device, and the system will automatically deduce the tensor sharding and communication patterns.
However, with the recent development in large-scale DL models, distributed training exhibits \textit{spatial} and \textit{temporal} workload heterogeneity, arising from both device disparities (e.g., mixed hardware, failures) and data variations (e.g., uneven sequence lengths).
Such heterogeneity violates SPMD's assumption of symmetric workload partitioning, which restricts its ability to express and optimize heterogeneous parallel strategies effectively. 

To address this, we propose HSPMD within the Hetu v2 system to achieve general and scalable DL training. 
HSPMD extends SPMD’s \textit{declarative annotations} to support asymmetric sharding and composes standard communication primitives for hierarchical communication, all while retaining the simplicity of a single-device programming model. HSPMD handles \textit{spatial} heterogeneity through progressive graph specialization, enabling device-specific execution logic, and addresses \textit{temporal} heterogeneity via dynamic graph switching. Evaluations on (a) heterogeneous devices, (b) unstable devices, and (c) mixed-length data scenarios show that HSPMD matches or outperforms specialized systems, providing a flexible and efficient solution for modern distributed DL training. Code is available: {\small \url{https://github.com/PKU-DAIR/Hetu}}.
\end{abstract}


\maketitle

\myvspace{-5pt}
\section{Introduction}
\label{sec:intro}

The rapid advancement of large-scale deep learning (DL) models, particularly large language models (LLMs)~\cite{ScalingLaws, db-gpt, llm-agents, llm-dataset, agent-rise, dl-code}, has pushed the boundaries of distributed training systems. Models like GPT~\cite{GPT4}, Gemini~\cite{Gemini}, and DeepSeek~\cite{DeepSeek-V3} routinely span billions of parameters, requiring a large number of devices working in concert for efficient training. 

To meet this demand, parallel strategies have evolved from singular approaches to sophisticated hybrid schemes combining data~\cite{PyTorchDistributed, PyTorchFSDP}, sequence~\cite{DeepSpeedSP, RingAttention, StripedAttention}, tensor~\cite{Megatron-LM, UsingMegatron-LM, MegatronSP}, and pipeline~\cite{GPipe, PipeDream, ZeroBubblePP, BPipe} parallelism. Within this landscape, the Single-Program Multiple-Data (SPMD) paradigm has emerged as a dominant abstraction due to its simplicity and scalability. By augmenting a single program with \textit{declarative annotations}, SPMD enables automatic parallelization while abstracting away low-level deployment details. For example, systems such as GSPMD~\cite{GSPMD} and Alpa~\cite{Alpa} annotate tensors with sharding semantics like \textit{Split} and \textit{Duplicate}, while DeepSpeed~\cite{Zero, DeepSpeedSP} and Megatron~\cite{Megatron-LM, UsingMegatron-LM, MegatronSP} annotate model layers with distributed variants (e.g., extending \textsf{Linear} into \textsf{ColumnParallelLinear} and \textsf{RowParallelLinear}). This approach decouples parallel strategy specification from the DL model, facilitating more scalable distributed training.

At the core of this declarative SPMD paradigm lies a fundamental premise: the training workload should be uniformly partitioned and distributed. This is because SPMD primitives are built to symmetrically shard tensors or layers across devices, following the assumptions that \textbf{(\lowercase\expandafter{\romannumeral1})} hardware resources are homogeneous and stable, and \textbf{(\lowercase\expandafter{\romannumeral2})} training burdens from all input data are equivalent. However, as DL training continues to scale, heterogeneous workloads have become increasingly prevalent. In practice, both hardware devices and input data introduce heterogeneity, as summarized in Table~\ref{tb:heterogeneity_causes}, leaving the standard SPMD design poorly suited to such scenarios. 

\mysubsubsection{Multifaceted heterogeneity} \textbf{(\lowercase\expandafter{\romannumeral1})} From the device perspective, heterogeneous and unstable devices are common in real-world cluster/cloud environments. On the one hand, due to the GPU shortage problem~\cite{SkyPilot, CloudGPUShortage}, modern MLaaS platforms typically host mixed GPU generations~\cite{MLaas, MultiTenantGPU}, and the scarcity of high-end GPUs (e.g., A100, H100) has led cloud providers to leverage heterogeneous GPUs with divergent capabilities (e.g., computation FLOPS, memory capacities, network bandwidth) for training~\cite{Sailor, cephalo}. On the other hand, device instability like GPU/node failures~\cite{MultiTenantGPU} occurs frequently. For example, Llama 3's training underwent 419 unexpected interruptions over 54 days, with 148 attributable to faulty GPUs~\cite{Llama3}. \textbf{(\lowercase\expandafter{\romannumeral2})} From the data perspective, the symmetric design of SPMD is a good fit when all input data are associated with equivalent workloads (e.g., preprocessed tabular datasets). 
However, to achieve general intelligence, modern DL models are often trained with unstructured, raw data, which demonstrates inherent length variations (e.g., varying text lengths~\cite{Packing}, diverse image resolutions~\cite{PatchPack}, differing video/audio durations~\cite{FasterVideo}). 

In short, these pervasive forms of heterogeneity fundamentally break the symmetry assumption of the standard SPMD paradigm and give rise to an urgent demand for asymmetric execution to effectively tackle heterogeneity.

\begin{figure}[!t]
\centering
\includegraphics[width=\linewidth]{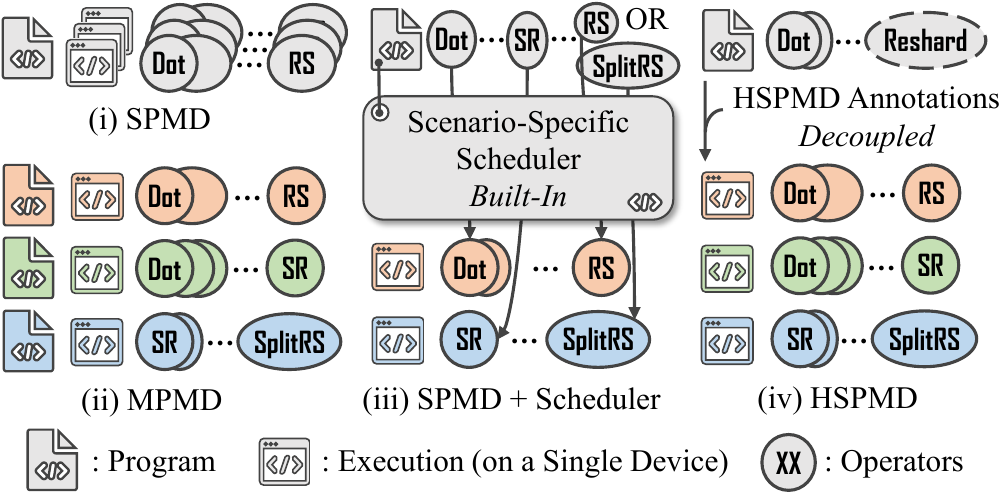}
\myvspace{-20pt}
\caption{\small{Comparison of SPMD, MPMD, scheduler-enhanced SPMD variants, and \system. While SPMD imposes strict symmetry, the others enable asymmetric execution to tackle heterogeneity.}}
\label{fig:xpmd}
\myvspace{-15pt}
\end{figure}

\mysubsubsection{Current solutions and limitations}
As illustrated in Figure~\ref{fig:xpmd} (left), one alternative is the Multiple-Program Multiple-Data (MPMD) paradigm, which uses distinct programs to encode different execution logic. However, MPMD often suffers from limited scalability and user-friendliness, as discussed in \S\ref{subsec:motivation}. Consequently, most distributed DL training systems addressing heterogeneity still adhere to the SPMD paradigm. As summarized in Table~\ref{tb:heterogeneity_causes}, a wide array of recent studies have extended standard SPMD to specific heterogeneous scenarios, including: heterogeneous devices~\cite{HyPar, HAP, Whale, AMP, HETHUB, Metis, HexiScale, Sailor}, unstable devices (i.e., elastic training)~\cite{Varuna, Bamboo, Oobleck, ReCycle}, and mixed-length data~\cite{HotSPa, FlexSP, ByteScale, WLB-LLM, Zeppelin, DCP}. A common approach among them is to integrate custom schedulers (e.g., HexiScale's heterogeneous pipeline scheduler~\cite{HexiScale}, Oobleck's elastic pipeline scheduler~\cite{Oobleck}) to enable asymmetric execution behaviors beyond standard SPMD, as shown in Figure~\ref{fig:xpmd} (middle). However, although these schedulers address heterogeneity in their respective scenarios, they are largely scenario-specific and built into the system with tight coupling to particular workloads. This design limits their flexibility and prevents them from serving as a general-purpose solution.

\begin{table}[!t]
\centering
\caption{\small{Example heterogeneous scenarios arising from device and data aspects, along with their characteristics and representative specialized SPMD training systems addressing each scenario.}}
\label{tb:heterogeneity_causes}
\myvspace{-8pt}
\small
\begin{tabular}{c|c|c|c}
\hline
\toprule
\multirow{2}{*}{Scenarios} & \multicolumn{2}{c|}{Characteristics} & \multirow{2}{*}{Representatives} \\
\cline{2-3}
 & Spatial & Temporal & \\
\midrule
\specialcell{Heterogeneous\\Devices} & \checkmark & & 
\specialcell{HexiScale~\cite{HexiScale}\\Sailor~\cite{Sailor}, etc.}
\\
\hline
\specialcell{Unstable\\ Devices} & \checkmark & \checkmark &
\specialcell{Oobleck~\cite{Oobleck},\\ReCycle~\cite{ReCycle}, etc.}
\\
\hline
\specialcell{Mixed-length\\Data} & \checkmark & \checkmark &
\specialcell{HotSPa~\cite{HotSPa},\\WLB-LLM~\cite{WLB-LLM}, etc.}
\\
\bottomrule
\hline
\end{tabular}
\myvspace{-15pt}
\end{table}

\mysubsubsection{Our solution and contributions}
Compared with prior case-by-case solutions, we propose \system (Hierarchical and Heterogeneous SPMD) within Hetu v2, a system that addresses multifaceted heterogeneity from a more general and fundamental perspective. \textbf{(\lowercase\expandafter{\romannumeral1}) Primitive-level extensions.} As shown in Figure~\ref{fig:xpmd} (right), instead of layering extensive scheduler-level efforts, \system pushes extensibility down to the primitive level of \textit{declarative annotations}. Specifically, we design the \system sharding annotations (\S\ref{sec:annotation}) and hierarchical communication resolution (\S\ref{sec:comm}), which enable asymmetric sharding and communication beyond SPMD’s inherent symmetry. While preserving SPMD’s core principle of separating the parallel strategy from a single programming view of the DL model, \system's primitives enable asymmetric execution natively, without the need for crafting scenario-specific schedulers. \textbf{(\lowercase\expandafter{\romannumeral2}) Characteristic-driven abstractions}. Departing from prior work that categorizes solutions by application scenarios (Table~\ref{tb:heterogeneity_causes}), we revisit and disentangle heterogeneity through its intrinsic characteristics: \textit{spatial} and \textit{temporal} (detailed in \S\ref{subsec:motivation}). \textit{Spatial} heterogeneity is due to the imbalanced workload and necessitates spatially heterogeneous parallel strategies (Figure~\ref{fig:three_scene_spatial}), while \textit{temporal} heterogeneity arises from dynamic workload changes and requires temporal reconfiguration across strategies (Figure~\ref{fig:two_scene_temporal}). Accordingly, we introduce graph specialization (\S\ref{sec:graph_specialization}) and graph switching (\S\ref{sec:graph_switching}), respectively. While these abstractions are scenario-agnostic individually, they serve as modular building blocks that can be composed into scenario-specific solutions, allowing \system to generalize beyond single-purpose SPMD variants.

In summary, this paper makes the following contributions.

\begin{itemize}[noitemsep, topsep=0pt, parsep=0pt, partopsep=0pt, leftmargin=*]
\item We introduce primitive-level extensions to break SPMD’s inherent symmetry, directly bridging single-device programming with heterogeneous parallelization.
\item We present a general system design methodology that builds upward from fundamental heterogeneity characteristics, rather than starting from specific heterogeneous scenarios and customizing downward as in prior specialized systems.
\item We evaluate \system across diverse training scenarios, covering (a) heterogeneous devices, (b) unstable devices, and (c) mixed-length data. Empirical results show that \system is at least on par with, and in most cases surpasses, state-of-the-art SPMD systems and specialized SPMD variants.
\end{itemize}

\myvspace{-5pt}
\section{Background and Motivations}
\label{sec:background}

\subsection{Sharding and Communication in Training}
\label{subsec:shard_and_comm}

As model sizes increase, the computational and memory resources of a single device become insufficient to support training, necessitating the adoption of distributed training. This introduces two critical challenges: \textbf{(\lowercase\expandafter{\romannumeral1})} how to effectively manage the \textbf{sharding} of different training components across devices; and \textbf{(\lowercase\expandafter{\romannumeral2})} how to efficiently coordinate their \textbf{communication}. To address these challenges, a variety of parallelisms and distributed training techniques have been proposed. 

\mysubsubsection{Data sharding} 
Data parallelism (DP)~\cite{PyTorchFSDP, PyTorchDistributed} shards the input data along the batch dimension, enabling different model replicas to process data simultaneously and synchronize gradients via all-reduce. For long sequences, techniques like sequence or context parallelism  (SP~\cite{DeepSpeedSP}  or CP~\cite{RingAttention, StripedAttention}) can be applied to further shard the sequence dimension. These require additional communication (e.g., all-to-all or ring-based send-receive) to ensure correctness of the attention~\cite{AttentionIsAllYouNeed}.

\mysubsubsection{Model sharding} Pipeline parallelism (PP)~\cite{GPipe, PipeDream, ZeroBubblePP, BPipe, pipeline-overview} shards model weights across layers, where activations are communicated between pipeline stages via send-receive. Besides, tensor parallelism (TP)~\cite{Megatron-LM, UsingMegatron-LM, MegatronSP} shards model weights along the hidden size dimension, requiring frequent all-gather and reduce-scatter on activations at each layer.

\mysubsubsection{Optimizer states sharding} 
Optimizer states consume significant memory. ZeRO~\cite{Zero} fully shards them across devices, necessitating all-gather and reduce-scatter operations, while some other methods~\cite{InternEvo, AMSP, Hydraulis} use finer-grained sharding, balancing storage redundancy and communication costs.

These techniques are often combined~\cite{Galvatron, Galvatron2, spindle, Malleus, lobra, Elastor}, creating complex sharding and communication patterns that challenge system-level expression and handling at scale.

\begin{figure}[!t]
\centering
\includegraphics[width=\linewidth]{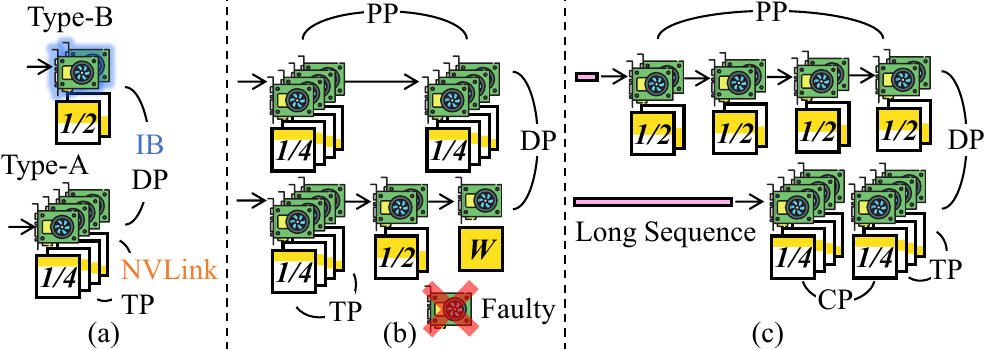}
\myvspace{-20pt}
\caption{\small{Use spatially heterogeneous parallel strategies to tackle spatial heterogeneity in scenario (a)-(c). ``DP, CP, TP, PP'' represent data, context, tensor, and pipeline parallelism, respectively. And scenarios (a), (b), and (c) refer to heterogeneous devices, unstable devices, and mixed-length data, respectively.}}
\label{fig:three_scene_spatial}
\myvspace{-15pt}
\end{figure}

\myvspace{-5pt}
\subsection{SPMD Training Systems}
\label{subsec:SPMD_training}

DL training can be modeled as a directed acyclic graph (DAG), where nodes represent operations and edges denote data dependencies. Tensors are the fundamental units flowing through the graph, carrying numerical values and metadata (e.g., shape, type). Operators transform input tensors into outputs (e.g., dot, attention). Following this, SPMD has emerged as a powerful paradigm (e.g.,  GSPMD~\cite{GSPMD}, Alpa~\cite{Alpa}, Unity~\cite{Unity}, DTensor~\cite{TensorFlowDTensor, PyTorchDTensor}). As shown in Figure~\ref{fig:SPMD_vs_HSPMD} (left), by specifying the DAG with \textit{declarative annotations}, SPMD decouples the programming from the complexities of parallelization, enabling automatic derivation and accommodation for diverse sharding and communication patterns. 

\mysubsubsection{Limitations}
Despite its advantages, the SPMD paradigm imposes strict symmetry constraints. First, existing SPMD annotations are confined to the uniform partitioning of tensors across regular device meshes. Additionally, SPMD-style collective communication primitives, such as all-reduce, all-gather, and reduce-scatter, require symmetric participation from devices. These two constraints make it challenging for SPMD to handle \textbf{asymmetric sharding} and \textbf{communication}.

\begin{figure}[!t]
\centering
\includegraphics[width=\linewidth]{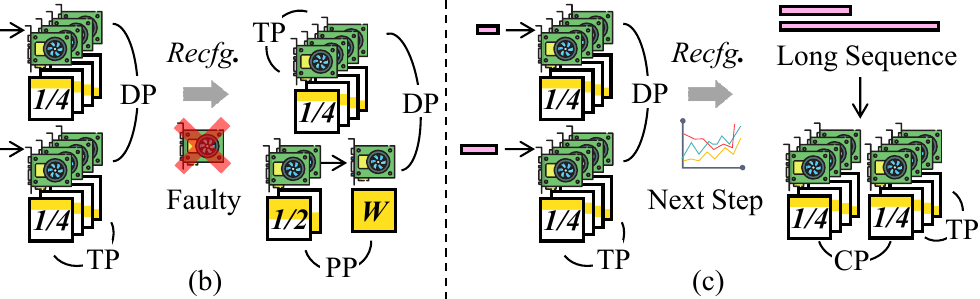}
\myvspace{-20pt}
\caption{\small{For scenarios (b) and (c), beyond the spatial heterogeneity shown in Figure~\ref{fig:three_scene_spatial}, heterogeneity may also evolve over time, demanding temporal reconfiguration between parallel strategies. (Scenario (a) is omitted because it exhibits only static spatial heterogeneity.)}}
\label{fig:two_scene_temporal}
\myvspace{-15pt}
\end{figure}

\myvspace{-5pt}
\subsection{Other Paradigms and Our Motivations}
\label{subsec:motivation}

As mentioned earlier, both devices and data can induce heterogeneity, leading to scenarios such as: (a) heterogeneous devices, (b) unstable devices, and (c) mixed-length data.

\mysubsubsection{MPMD vs. SPMD}
Due to its symmetry constraints, SPMD is ill-suited for these diverse heterogeneous scenarios. 
As shown in Figure~\ref{fig:xpmd} (left), one alternative is the MPMD paradigm (e.g., Pathways~\cite{pathways}, Ray~\cite{Ray}). However, MPMD is more suited for multi-task settings (e.g., RL for LLMs~\cite{verl, staleflow}), where heterogeneity manifests across tasks. By contrast, in single-task training with intra-task heterogeneity, MPMD typically requires generating and managing thousands of distinct programs across a large cluster, incurring substantial compilation overheads~\cite{GShard} and leading to a poor user experience. As a result, the scalability and simplicity of the single-program paradigm in SPMD remain highly desirable at large scale. 

\mysubsubsection{Scheduler-enhanced SPMD variants}
Consequently, many systems continue to build upon SPMD but integrate a high-level scheduler to enable heterogeneous execution logic, as shown in Figure~\ref{fig:xpmd} (middle). While maintaining a single programming model, this approach requires workarounds to bypass SPMD’s inherent symmetry, which introduces extensive control and branching logic within the scheduler. More critically, these schedulers are built-in and scenario-specific, making such SPMD variants only applicable to a single scenario.

\mysubsubsection{Motivations}
This motivates a more principled approach beyond scheduler-level customization. Specifically, we ask: can SPMD itself provide intrinsic support for asymmetry? A natural direction is to extend its \textit{declarative annotations}. Compared with built-in schedulers, this approach offers better decoupling, natively expresses asymmetric parallelization, and provides a unified substrate across diverse scenarios.

Moreover, prior work treats each heterogeneous scenario as an isolated case,  requiring dedicated system mechanisms and ad hoc extensions, which is fundamentally unscalable. This motivates us to seek a more essential characterization that cuts across diverse scenarios. Specifically, we distill two fundamental dimensions underlying multifaceted heterogeneity: \textit{spatial} heterogeneity and \textit{temporal} heterogeneity (Table~\ref{tb:heterogeneity_causes}).

\textit{Spatial} heterogeneity refers to workload imbalance at a given moment, calling for spatially heterogeneous parallel strategies. For example (Figure~\ref{fig:three_scene_spatial}): (a) for heterogeneous devices, one can adopt a higher TP for memory-constrained GPUs and a lower TP for others; (b) for unstable devices, heterogeneous TP grouping and PP composition allow the remaining GPUs (e.g., 15 GPUs) to be fully utilized when one GPU fails; (c) for mixed-length data, GPUs can be partitioned into subgroups optimized for distinct sequence lengths (e.g., employing larger TP or CP for long sequences). 

\textit{Temporal} heterogeneity reflects workload variation over time, necessitating dynamic reconfiguration of parallel strategies. For example (Figure~\ref{fig:two_scene_temporal}): (b) for unstable devices, when a GPU/node becomes unavailable, the overall parallel strategy must be reconfigured; (c) for mixed-length data, when the distribution of sequence lengths shifts across training steps (e.g., a step with very long sequences appears), reconfiguring to a strategy tailored for long sequences is essential to ensure better performance or to avoid out-of-memory (OOM) errors.

\begin{figure}[!t]
\centering
\includegraphics[width=\linewidth]{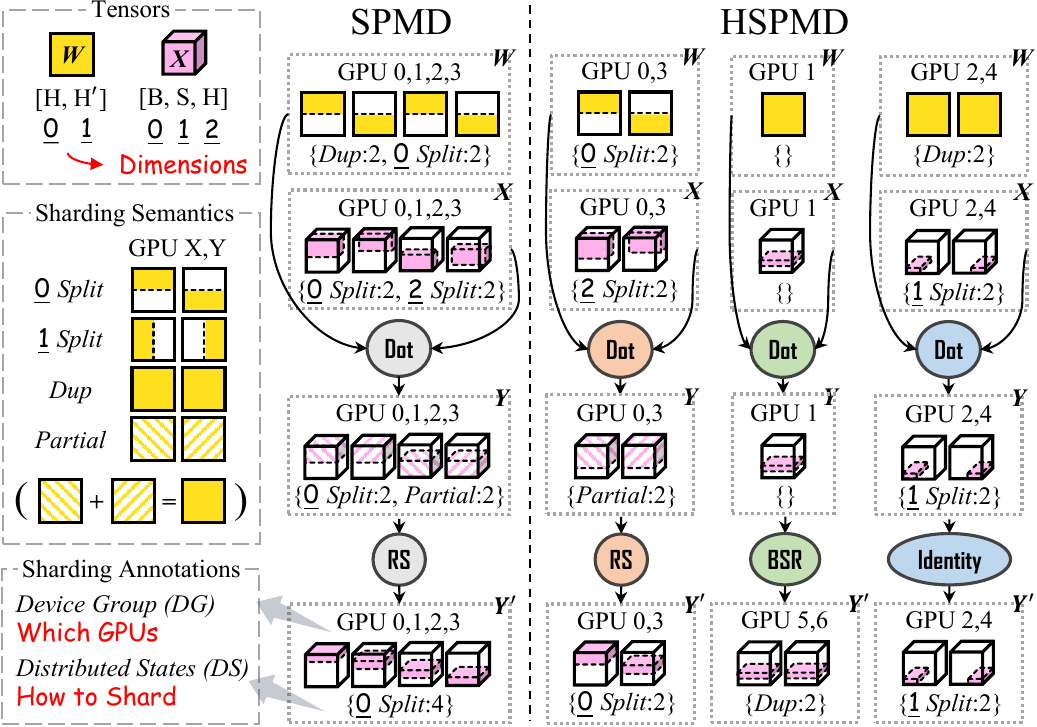}
\myvspace{-20pt}
\caption{\small{Left: Standard SPMD-style sharding annotations (e.g., DTensor~\cite{TensorFlowDTensor, PyTorchDTensor}) and their deduction when employing DP and TP across GPU 0-3. They can only express symmetric sharding and communication. Right: An example of \system expressing asymmetry: TP between GPU 0,3 and GPU 5,6; PP between GPU 1 and GPU 5,6; CP between GPU 2,4; and DP across them. ``\texttt{RS}, \texttt{BSR}'' represents reduce-scatter and batched-send-receive (\S\ref{subsec:bsr}), respectively.}}
\label{fig:SPMD_vs_HSPMD}
\myvspace{-15pt}
\end{figure}

To this end, we introduce \system within Hetu v2. \textbf{(\lowercase\expandafter{\romannumeral1})} At the primitive level, as illustrated in Figure~\ref{fig:SPMD_vs_HSPMD} (right), \system extends SPMD to natively support asymmetric sharding (\S\ref{sec:annotation}) and communication (\S\ref{sec:comm}), while preserving single-program semantics and collective operations. \textbf{(\lowercase\expandafter{\romannumeral2})} At the abstraction level, to address \textit{spatial} and \textit{temporal} heterogeneity, we propose graph specialization (\S\ref{sec:graph_specialization}) and graph switching (\S\ref{sec:graph_switching}), enabling heterogeneous deployment and dynamic reconfiguration, respectively. Together, these innovations establish \system as a general system for multifaceted heterogeneity.

\myvspace{-5pt}
\section{\system Design and Overview}
\label{sec:workflow}

\begin{figure*}[!t]
\centering
\includegraphics[width=\linewidth]{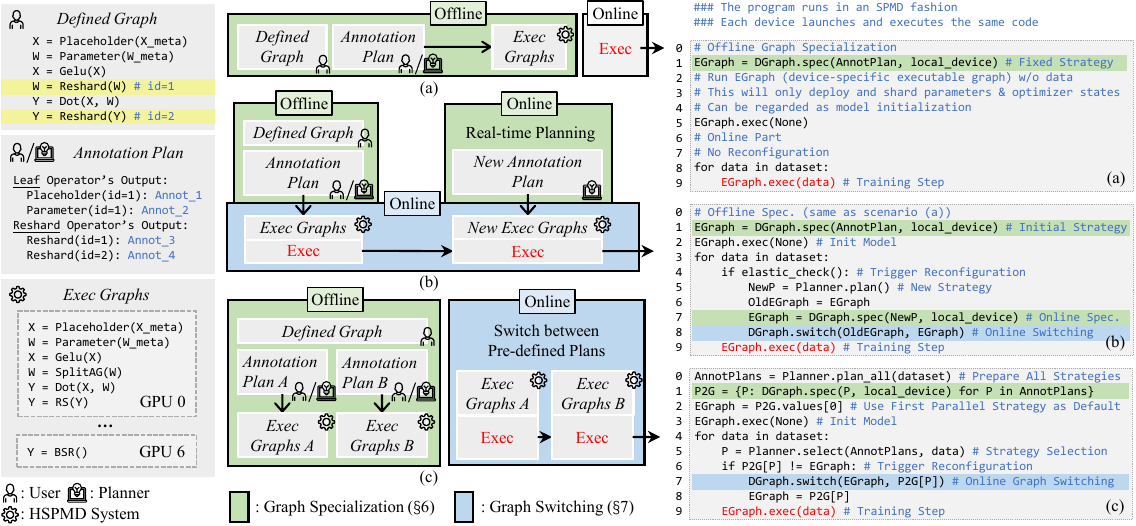}
\myvspace{-20pt}
\caption{\small{\system's workflow and code snippets (DL operators align with Figure~\ref{fig:SPMD_vs_HSPMD}). Our system relies on two abstractions: graph specialization (\S\ref{sec:graph_specialization}), which generates device-specific \textit{executable graphs} from a \textit{defined graph} (provided by users) and an \textit{annotation plan} (provided by users or planners); and graph switching (\S\ref{sec:graph_switching}), which enables switching between arbitrary \textit{executable graphs}.
These two abstractions act as building blocks that can be flexibly applied to: (a) heterogeneous devices, (b) unstable devices, and (c) mixed-length data.}}
\label{fig:workflow}
\myvspace{-15pt}
\end{figure*}

In this section, we provide a walkthrough of \system's core design and workflow.
As depicted in Figure~\ref{fig:workflow} (left), following prior work~\cite{PyTorchDistributed, FlexFlow, Alpa}, \system adopts a graph-based representation.
Users provide a single \textit{defined graph} consisting of three types of operators:
\textbf{(\lowercase\expandafter{\romannumeral1})} Leaf operators (e.g., placeholders, parameters) that produce data, model parameters, or optimizer states;
\textbf{(\lowercase\expandafter{\romannumeral2})} \texttt{Reshard} operators, introduced by \system as abstract markers to indicate potential resharding of intermediate results (e.g., gathering activations, resharding half-precision parameters);
\textbf{(\lowercase\expandafter{\romannumeral3})} All other operators that encode the model’s computation logic. An \textit{annotation plan}, either user-specified or generated by an external planner, is then associated with the \textit{defined graph}. This plan assigns output sharding annotations (\S\ref{sec:annotation}) to \textbf{(\lowercase\expandafter{\romannumeral1})} leaf and \textbf{(\lowercase\expandafter{\romannumeral2})} \texttt{Reshard} operators, defining the initial sharding layout and resharding behavior of critical tensors, thereby collectively specifying the parallel strategy.

Given the \textit{defined graph} and the \textit{annotation plan}, \system deduces the output sharding of \textbf{(\lowercase\expandafter{\romannumeral3})} all other operators and resolves how each \texttt{Reshard} operator can be concretely achieved by communication operators (e.g., reduce-scatter, all-gather) automatically (\S\ref{sec:comm}).\footnote{Not all communication needs to be explicitly represented via \texttt{Reshard}. For example, CP involves ring-based send–receive within attention~\cite{RingAttention}. We provide a fused \texttt{ParallelAttn} operator to encapsulate such implicit communication, keeping it transparent to the annotation system.} 
Additionally, \system will transparently handle the deployment and execution of parallel strategies, allowing users to write a single program by simply providing the \textit{defined graph} (and optionally the \textit{annotation plan}), without engaging in manual orchestration of distributed training.

\system leverages two abstractions to tackle heterogeneity (Figure~\ref{fig:workflow}, middle). 
Firstly, to address \textit{spatial} heterogeneity, where devices may require divergent execution logic, we introduce graph specialization (\S\ref{sec:graph_specialization}). Once the \textit{defined graph} and \textit{annotation plan} are in place, each device specializes its local operators, generating a device-specific \textit{executable graph} for deployment and execution. 
Secondly, to address \textit{temporal} heterogeneity, where the parallel strategy needs to be reconfigured, we introduce graph switching (\S\ref{sec:graph_switching}). When a new \textit{annotation plan} (e.g., changes in parameter sharding) is introduced, \system facilitates seamless switching between parallel strategies by online resharding the necessary weights between \textit{executable graphs}, without requiring a restart.

As illustrated in Figure~\ref{fig:workflow} (middle and right), these two abstractions, graph specialization and switching, are modular building blocks that can be flexibly combined to accommodate diverse heterogeneous scenarios. 
\textbf{(a) Heterogeneous devices} (\textit{spatial} heterogeneity only): An expert user or scenario-specific planner provides an \textit{annotation plan} specifying a heterogeneous parallel strategy. \textit{Executable graphs} are specialized once offline (line 1) and remain fixed throughout training (line 9).
\textbf{(b) Unstable devices} (\textit{spatial} + unknown \textit{temporal} heterogeneity): The initial setup mirrors case (a). Upon triggering a reconfiguration (e.g., due to changes in device availability), an automatic planner generates a new \textit{annotation plan} (line 5), and the system specializes new \textit{executable graphs} at runtime (line 7). Graph switching then seamlessly transitions model and optimizer states (line 8), ensuring uninterrupted training (line 9).
\textbf{(c) Mixed-length data} (\textit{spatial} + predictable \textit{temporal} heterogeneity): Users or planners prepare multiple \textit{annotation plans} (line 0), each optimized for a particular sequence length distribution, with corresponding \textit{executable graphs} pre-specialized offline (line 1). During training, the system dynamically selects and switches to the optimal plan when input distribution shifts (line 5-8).

Although scenario-specific planners are needed (e.g., for \textit{annotation plan} generation and selection), \system is designed as a general framework to address diverse forms of heterogeneity, rather than to optimize automatic planning for a single scenario. Our focus is on introducing system core designs, while keeping the planner modular and easily replaceable. 
In the following, we first detail the sharding annotations (\S\ref{sec:annotation}) and communication resolution (\S\ref{sec:comm}), which facilitate asymmetric sharding and communication, respectively. We then show how \system addresses \textit{spatial} heterogeneity with graph specialization (\S\ref{sec:graph_specialization}) and \textit{temporal} heterogeneity via graph switching (\S\ref{sec:graph_switching}). Finally, we briefly introduce \system's planners (\S\ref{sec:impl}).

\myvspace{-5pt}
\section{Sharding Annotations}
\label{sec:annotation}

This section introduces \system's sharding annotations, which hierarchically extend SPMD-style annotations through a bottom-up approach to support asymmetric sharding. Compared to crafting high-level specialized schedulers on top of SPMD, this low-level extension enables SPMD itself to express asymmetry and generalize to diverse scenarios.

\myvspace{-5pt}
\subsection{Basic SPMD-style Annotations}
\label{subsec:basic-spmd-annotations}

Following existing SPMD-style \textit{declarative annotations} such as GSPMD~\cite{GSPMD} and DTensor~\cite{TensorFlowDTensor, PyTorchDTensor}, we associate each tensor with two attributes: \textbf{(\lowercase\expandafter{\romannumeral1})} \textbf{\textit{Device Group (DG)}}, which specifies the devices where the tensor resides; \textbf{(\lowercase\expandafter{\romannumeral2})} \textbf{\textit{Distributed States (DS)}}, which defines how the tensor is sharded across these devices. While some critical tensors (e.g., data, parameters, \texttt{Reshard} outputs) have their \DeviceGroup and \DistributedStates explicitly specified by the \textit{annotation plan}, others are deduced automatically. 

As shown in Figure~\ref{fig:SPMD_vs_HSPMD} (left), the \DeviceGroup is represented as an ordered list of GPU indices, and the \DistributedStates is an ordered dictionary where the key $dim$ corresponds to a logical distributed dimension (a virtual axis for sharding), and its corresponding value represents the number of shards along $dim$. Following previous works, there are three sharding semantics: \textbf{(\lowercase\expandafter{\romannumeral1})} \textbf{\textit{Split}} ($dim \geq 0$), where the tensor is uniformly split along its physical shape dimension $dim$; \textbf{(\lowercase\expandafter{\romannumeral2})}  \textbf{\textit{Duplicate}}  ($dim = -1$), where the tensor is fully replicated; \textbf{(\lowercase\expandafter{\romannumeral3})} \textbf{\textit{Partial}} ($dim = -2$), where the tensor's values are partially stored. These annotations come with inherent deduction rules. For instance in Figure~\ref{fig:SPMD_vs_HSPMD} (left), given the annotations of $W$ and $X$, the annotation of their \texttt{Dot} result can be derived: $Y$ retains the same \DeviceGroup as the inputs, while its \DistributedStates transforms the original \textit{split} into \textit{partial}.

\subsection{Hierarchical \& Heterogeneous Extensions}
\label{subsec:extended-annotations }

Building on basic SPMD annotations, we introduce hierarchical constructs to support asymmetric sharding, enabling heterogeneous execution logic as shown in Figure~\ref{fig:SPMD_vs_HSPMD} (right).

\begin{figure}[!t]
\centering
\includegraphics[width=\linewidth]{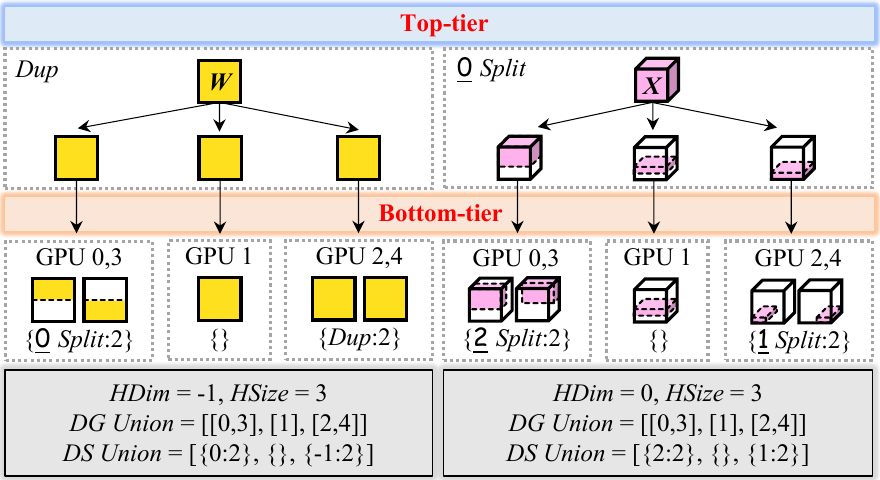}
\myvspace{-20pt}
\caption{\small{\system's sharding annotations (aligned with Figure~\ref{fig:SPMD_vs_HSPMD}). The tensor is first sharded along the \HeteroDim dimension into \HeteroSize \textit{sharding subgroups}. The devices in each \textit{sharding subgroup} are given by the corresponding \DeviceGroup within the \DeviceGroupUnion. And each \textit{sharding subgroup} applies a \DistributedStates from the \DistributedStatesUnion. In this example, both $W$ and $X$ contain 3 \textit{sharding subgroups} (\HeteroSize = 3).}}
\label{fig:annotations}
\myvspace{-10pt}
\end{figure}

\textbf{(\uppercase\expandafter{\romannumeral1}) \textit{Device Group Union (DG Union)} and \textit{Distributed States Union (DS Union)}.} As illustrated in Figure~\ref{fig:annotations}, we generalize the single \DeviceGroup\ and \DistributedStates\ associated with each tensor to multiple \DeviceGroup\ partitions and \DistributedStates\ configurations, forming \DeviceGroupUnion\ and \DistributedStatesUnion. Each \DistributedStates\ configuration in the union corresponds to a \DeviceGroup\ partition in the union, describing sharding within that particular device subgroup, referred to as a \textit{sharding subgroup}.\footnote{\textit{Sharding subgroups} must consist of mutually exclusive device subsets.} We term the sharding within each subgroup as \textbf{bottom-tier sharding}. This approach enables the simultaneous representation of multiple sharding patterns.

\textbf{(\uppercase\expandafter{\romannumeral2}) \textit{Heterogeneous Dimension (HDim)} and \textit{Heterogeneous Size (HSize)}.} To model relationships between \textit{sharding subgroups} in a union, we further introduce \textit{Heterogeneous Dimension (HDim)}\ and \textit{Heterogeneous Size (HSize)}. As illustrated in Figure~\ref{fig:annotations}, \HeteroSize\ specifies the number of distinct \textit{sharding subgroups}, while \HeteroDim\ describes the sharding across them. For instance, $\HeteroDim = 0$ for tensor $X$ implies splitting the tensor's first dimension across subgroups, while $\HeteroDim = -1$ for tensor $W$ indicates full replication across subgroups. We refer to this as the \textbf{top-tier sharding}.\footnote{The shape of each shard can vary and is determined at runtime, which is decoupled from the static annotation system (as elaborated in \S\ref{subsec:exec}).}

\myvspace{-5pt}
\subsection{Discussions}
\label{subsec:annotation_discussions}

\system's annotations can be viewed as a sharding extension layered atop standard SPMD annotations. 
Although this extension introduces additional system complexity, much of it remains hidden from the users' view. For one thing, as introduced in \S\ref{sec:workflow}, users only need to define a logical computation graph (\textit{defined graph}) and optionally provide a small set of annotations (\textit{annotation plan}), the rest can be automatically deduced by our system. For another, \system's annotation deduction can be decomposed into multiple standard SPMD deductions within each \textit{sharding subgroup}, thereby simplifying the deduction procedure (this will be detailed in \S\ref{subsec:anno_deduc}).

Readers might wonder whether it is more preferable to consider a more granular multi-tier hierarchy, or a simpler scheme that only shards data unevenly across devices. We discuss these two alternatives below.
\begin{itemize}[noitemsep, topsep=0pt, parsep=0pt, partopsep=0pt, leftmargin=*]
\item \textit{Why two-tier, not multi-tier:} Introducing a multi-tier hierarchy is unnecessary. \HeteroDim and \HeteroSize at the top tier already suffice to express complex asymmetric shardings while keeping the added complexity minimal. Moreover, the two-tier design matches how compute resources are organized in practice: a node or node group is internally load-balanced under SPMD, while heterogeneity exists across nodes or node groups. Adding further tiers would enlarge the planning space and deduction complexity without yielding additional expressiveness for today's cluster deployment.

\item \textit{Why not simply shard data unevenly:} Sharding data alone, without a fundamental change at the annotation level, is insufficient. 
\textbf{(\lowercase\expandafter{\romannumeral1})} For \textit{temporal} heterogeneity, it is ineffective: when a GPU fails (Figure~\ref{fig:two_scene_temporal}(b)), the TP/PP degrees must be reconfigured, which cannot be handled by data re-sharding alone.
\textbf{(\lowercase\expandafter{\romannumeral2})} For \textit{spatial} heterogeneity, it is suboptimal: under low inter-cluster bandwidth, heterogeneous DP requires exchanging full gradients, whereas heterogeneous PP only transfers activations for a single layer, with much smaller volume. Capturing such PP strategies requires annotation-level support beyond simple data sharding.
\end{itemize}

Once tensor annotations are specified by the \textit{annotation plan} or deduced by the system, \system will determine the exact communication pattern for each \texttt{Reshard} operator and will instantiate it with device-specific communication operators that can be executed directly. The next section (\S\ref{sec:comm}) will describe how the asymmetric communication operators (e.g., \texttt{BSR} in Figure~\ref{fig:SPMD_vs_HSPMD},\ref{fig:workflow}) are derived from the given annotations.

\myvspace{-5pt}
\section{Hierarchical Communication Resolution}
\label{sec:comm}

\begin{figure}[!t]
\centering
\includegraphics[width=\linewidth]{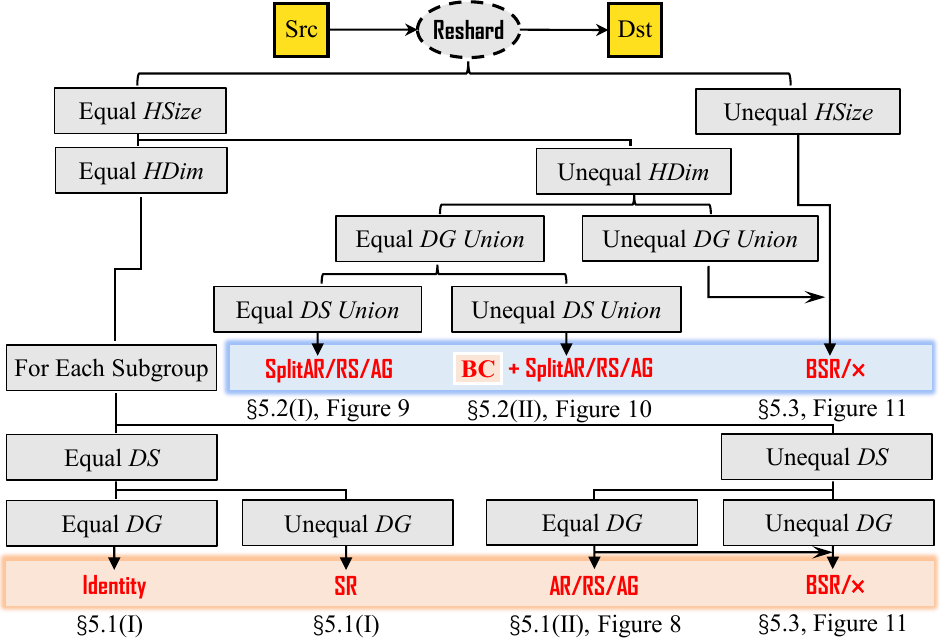}
\myvspace{-20pt}
\caption{\small{HSPMD's communication resolution. 
Orange denotes bottom-tier communication, executed independently within \textit{sharding subgroups}; blue indicates top-tier communication, involving interactions between \textit{sharding subgroups}. ``\texttt{SR}'': send-receive; ``\texttt{AR}, \texttt{RS}, \texttt{AG}'': all-reduce, reduce-scatter, all-gather; ``\texttt{BSR}'': batched-send-receive; ``$\times$'': unsupported; ``BC'': bottom-tier communication.}}
\label{fig:comm}
\myvspace{-10pt}
\end{figure}

As depicted in Figure~\ref{fig:comm}, to reshard from a source annotation to a destination one, we develop a comprehensive classification procedure to determine the appropriate communication operators. Aligning with the concepts of bottom-tier and top-tier sharding, we classify communication into two categories: \textbf{bottom-tier communication} and \textbf{top-tier communication}. Bottom-tier communication operates independently within each \textit{sharding subgroup}, determined solely by changes in bottom-tier sharding (\S\ref{subsec:bottom_comm}). In contrast, top-tier communication requires interactions across \textit{sharding subgroups} and is influenced by the entire annotation hierarchy (\S\ref{subsec:top_comm}).

During this procedure, we prioritize the use of collective communication operators, decomposing asymmetric communication into multiple symmetric collective communications when possible. For one thing, collective communication libraries (CCLs) have been extensively optimized~\cite{NCCL, TCCL, HiCCL, Swing}, often achieving higher performance than hand-crafted kernels. For another, modern GPU cluster network topologies (e.g., Multi-rail~\cite{TofuD, ThetaGPU}, Clos~\cite{ClosNetwork}) are typically organized into multiple tiers~\cite{Enfabrica}, which aligns well with our hierarchical design. Specifically, \textbf{(\lowercase\expandafter{\romannumeral1})} bottom-tier communication within \textit{sharding subgroups} is usually operated on homogeneous devices (intra-machine or inter-machine with uniform links, typically high bandwidth) and can exploit fast symmetric collective primitives, whereas \textbf{(\lowercase\expandafter{\romannumeral2})} top-tier communication handles potentially heterogeneous inter-subgroup links (e.g., IB/TCP mixtures, often slower). When collective communication is infeasible, we develop a batched-send-receive (\texttt{BSR}) operator (\S\ref{subsec:bsr}) to accomplish sophisticated communications.

\myvspace{-5pt}
\subsection{Bottom-tier Communication}
\label{subsec:bottom_comm}

When the source and destination annotations share the same \HeteroSize and \HeteroDim, it implies that the top-tier sharding remains unchanged. Consequently, all \textit{sharding subgroups} can individually perform communication, typically reducing to standard SPMD resolution. As illustrated in Figure~\ref{fig:comm}, for each \textit{sharding subgroup}, we consider the following cases:

\textbf{(\uppercase\expandafter{\romannumeral1}) Subgroup's source \DistributedStates equals destination \DistributedStates.} 
We then determine whether the \DeviceGroup changes from the source to the destination, in which case each corresponding device executes either an \texttt{Identity} or send-receive (\texttt{SR}) operator to directly transfer the local shard.

\begin{figure}[!t]
\centering
\includegraphics[width=\linewidth]{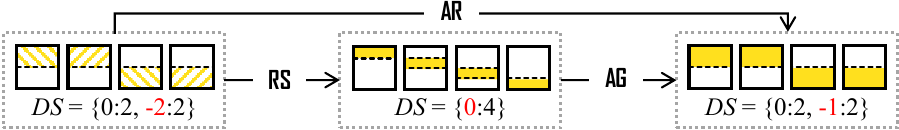}
\myvspace{-20pt}
\caption{\small{Bottom-tier collective communication.}}
\label{fig:bottom_comm}
\myvspace{-10pt}
\end{figure}

\textbf{(\uppercase\expandafter{\romannumeral2}) Subgroup's source \DistributedStates does not equal destination  \DistributedStates.} 
If the source and destination share the same \DeviceGroup, the device composition of the subgroup is unchanged, and only intra-subgroup resharding is required. As shown in Figure~\ref{fig:bottom_comm}, we analyze the \DistributedStates to select the appropriate collective operator, e.g., all-reduce (\texttt{AR}), reduce-scatter (\texttt{RS}), or all-gather (\texttt{AG}).

If these collective communication conditions are unmet, or if the \DeviceGroup differs, the system will instead employ batched-send-receive (\texttt{BSR}). Its mechanism will be detailed in \S\ref{subsec:bsr}.

\begin{figure}[!t]
\centering
\includegraphics[width=\linewidth]{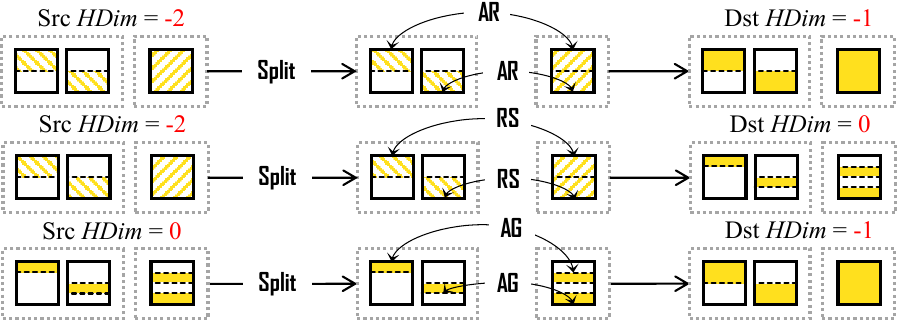}
\myvspace{-20pt}
\caption{\small{Top-tier collective communication. In scenarios where only \HeteroDim is altered, each shard can be split to the finest granularity, followed by multiple collective communication operators.}}
\label{fig:top_comm}
\myvspace{-10pt}
\end{figure}

\myvspace{-5pt}
\subsection{Top-tier Communication}
\label{subsec:top_comm}

Next, we examine the scenario where the \HeteroSize of the source matches that of the destination, but their \HeteroDim values differ. If the \DeviceGroupUnion of the source is identical to that of the destination (i.e., every \DeviceGroup in the union is equivalent), the transformation between the source and the destination can still be efficiently realized using collective communication operators. We analyze the following two cases:

\textbf{(\uppercase\expandafter{\romannumeral1}) Source \DistributedStatesUnion equals destination \DistributedStatesUnion.}  In this case, only the \HeteroDim changes between the source and destination. As depicted in Figure~\ref{fig:top_comm}, the transformation can be achieved through three distinct operators: split-all-reduce (\texttt{SplitAR}), split-reduce-scatter (\texttt{SplitRS}), and split-all-gather (\texttt{SplitAG}). The process starts by identifying the finest-grained slices of all \textit{sharding subgroups}. Subsequently, based on the \HeteroDim change, collective communication is performed across these \textit{sharding subgroups} for each slice, while maintaining the bottom-tier sharding within each subgroup. 

\begin{figure}[!t]
\centering
\includegraphics[width=\linewidth]{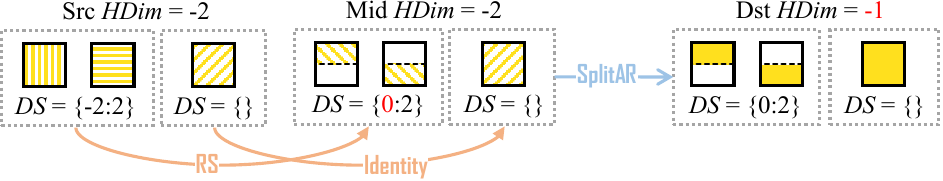}
\myvspace{-22pt}
\caption{\small{Top-tier collective communication succeeded by bottom-tier communication. In scenarios where both \HeteroDim and \DistributedStatesUnion are modified, we first align the \DistributedStates within each \textit{sharding subgroup}, followed by aligning the \HeteroDim across different \textit{sharding subgroups}.}}
\label{fig:two_comm}
\myvspace{-10pt}
\end{figure}

\textbf{(\uppercase\expandafter{\romannumeral2}) Source \DistributedStatesUnion does not equal destination \DistributedStatesUnion.}  
As illustrated in Figure~\ref{fig:two_comm}, in this case we first use bottom-tier communication (e.g., \texttt{RS}, \texttt{Identity}) to align the \DistributedStatesUnion, reducing the problem to (\uppercase\expandafter{\romannumeral1}). We then apply top-tier communication (e.g., \texttt{SplitAG}) to align the \HeteroDim.

For scenarios where the \textit{DG Unions} of the source and the destination are not identical, or where the \HeteroSize values differ, we once again resort to the \texttt{BSR} operator (\S\ref{subsec:bsr}).

\myvspace{-5pt}
\subsection{Batched-Send-Receive Mechanism}
\label{subsec:bsr}

\begin{figure}[!t]
\centering
\includegraphics[width=\linewidth]{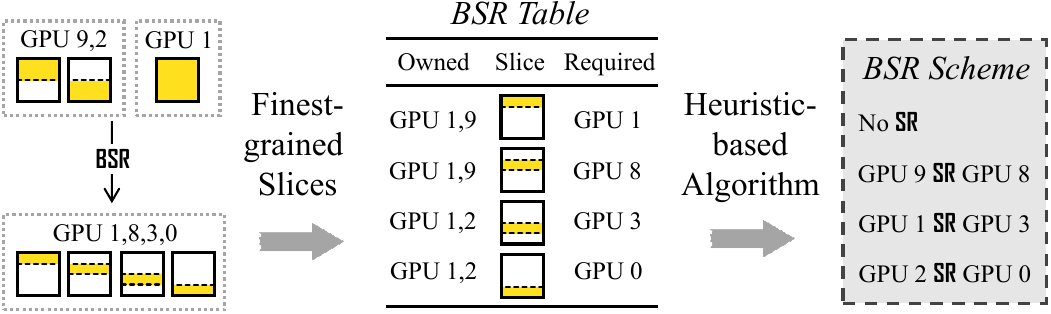}
\myvspace{-20pt}
\caption{\small{Batched-Send-Receive (\texttt{BSR}) mechanism. Based on the annotations of the source and destination, we construct a \textit{BSR table} and subsequently generate the \textit{BSR scheme} for communication.}}
\label{fig:bsr}
\myvspace{-10pt}
\end{figure}

When it does not involve \textit{Partial}, any complex resharding can be decomposed into multiple send and receive operators, referred to as batched-send-receive (\texttt{BSR}), which is particularly useful for handling complex communication patterns that cannot be supported by collective communications. 

As depicted in Figure~\ref{fig:bsr} (left), we begin by identifying the finest-grained slices, based on which we construct a \textit{BSR table} that maps each slice to its owner devices and the devices that require it. 
Subsequently, a \textit{BSR scheme} is generated to describe how the slices are transmitted. 
Particularly, as illustrated in Figure~\ref{fig:bsr} (right), we generate the \textit{BSR scheme} by sequentially scanning the \textit{BSR table} and applying the following heuristics: \textbf{(\lowercase\expandafter{\romannumeral1})} If the slice is already owned, we perform a direct copy without communication; \textbf{(\lowercase\expandafter{\romannumeral2})} We prefer links with higher bandwidth (e.g., GPU 9 sending to GPU 8 via NVLink); \textbf{(\lowercase\expandafter{\romannumeral3})} If bandwidths are equal, we balance the cumulative send load (e.g., GPU 1 and GPU 2 send alternately). 

\mysubsubsection{Discussions}
Since each slice may have multiple senders (owner devices) and receivers (required devices), deriving the optimal \textit{BSR scheme} necessitates solving a Generalized Assignment Problem (GAP)~\cite{GAProblem}, which is NP-hard. Our heuristic approach reduces the solving complexity to $O(pq)$, where $p$ is the number of entries in the \textit{BSR table} and $q$ is the maximum number of receivers per entry. Here, $p$ corresponds to the partition granularity of the finest-grained slices (e.g., $p = 4$ in Figure~\ref{fig:bsr}) and does not increase with cluster scale, since scaling is typically achieved by replication along the DP dimension rather than finer per-tensor partitioning. As a result, the solver remains tractable even at a large scale.

Our solver models only P2P bandwidth, which may be insufficient under highly heterogeneous network topologies. In such cases, a single P2P transfer may traverse multiple physical links, and the true bottleneck can reside in a specific link along the path. A more fine-grained model that captures per-link characteristics could better expose such intra-path bottlenecks. However, in practice, the \texttt{BSR} overhead is already small relative to computation (see \S\ref{sec:case_study}), suggesting that further optimizing the solver would yield limited benefits. Overall, though our heuristic-based algorithm may not achieve the optimal, it is more practical and easier to implement, with strong performance and low overhead (see \S\ref{sec:case_study}).

Last but not least, \texttt{BSR} is infeasible when \textit{Partial} is involved. 
However, in practice, \textit{Partial} tensors are typically intermediate results associated with collective communication, which can be handled by other top-tier or bottom-tier communication mechanisms. Given that complex resharding for \textit{Partial} tensors is generally unnecessary, we omit such scenarios, yet our work can be easily extended by integrating \texttt{BSR} with all-reduce (when source involves \textit{Partial}) or numeric division (when destination involves \textit{Partial}).

\myvspace{-5pt}
\section{Progressive Graph Specialization}
\label{sec:graph_specialization}

\begin{figure}[!t]
\centering
\includegraphics[width=\linewidth]{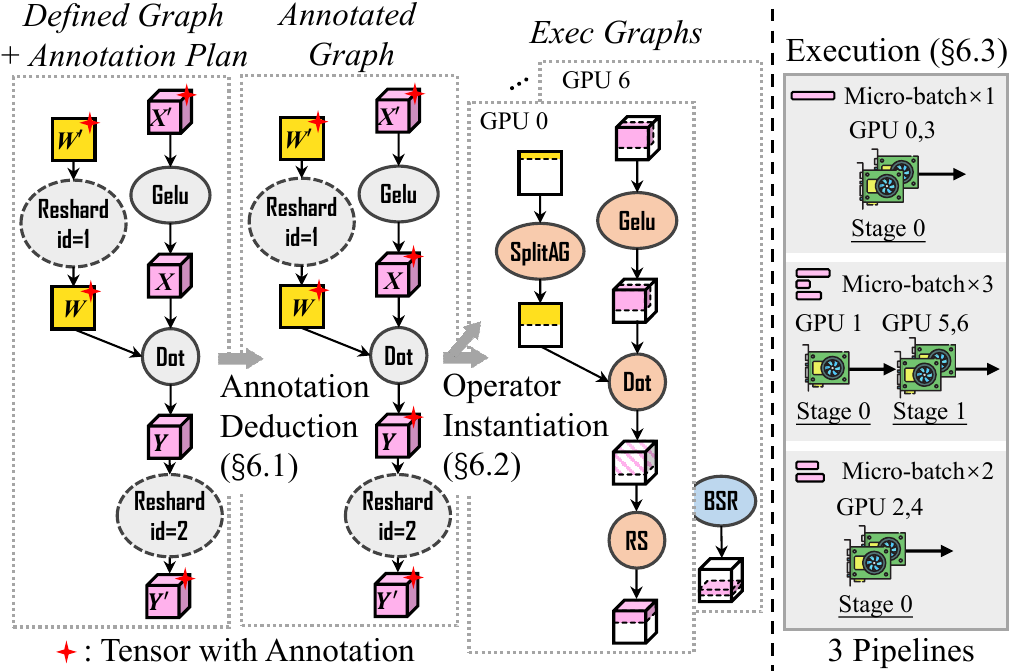}
\myvspace{-20pt}
\caption{\small{\system’s graph specialization. We systematically deduce and parse annotations from a unified \textit{defined graph}, transforming it into device-specific execution logic to support heterogeneous parallelization. The annotations remain consistent with Figure~\ref{fig:SPMD_vs_HSPMD}.}}
\label{fig:graph_specialization}
\myvspace{-10pt}
\end{figure}

In this section, we detail the abstraction of graph specialization that tackles \textit{spatial} heterogeneity. As shown in Figure~\ref{fig:graph_specialization}, the process starts from a \textit{defined graph} with initial annotations specified by the \textit{annotation plan} (i.e., the output sharding of leaf and \texttt{Reshard} operators). We then propagate these annotations to all remaining tensors, resulting in a fully \textit{annotated graph} (\S\ref{subsec:anno_deduc}). Next, we resolve the concrete communication of \texttt{Reshard} (as detailed in \S\ref{sec:comm}) and instantiate device-specific operators to generate distinct \textit{executable graphs} for each device (\S\ref{subsec:ops_inst}), which are then deployed for pipelined execution (\S\ref{subsec:exec}). Starting from a single-program abstraction, this specialization process incrementally derives divergent execution logics across devices, enabling the expression of heterogeneous parallel strategies that address \textit{spatial} heterogeneity.

\myvspace{-5pt}
\subsection{Annotation Deduction}
\label{subsec:anno_deduc}

\begin{figure}[!t]
\centering
\includegraphics[width=\linewidth]{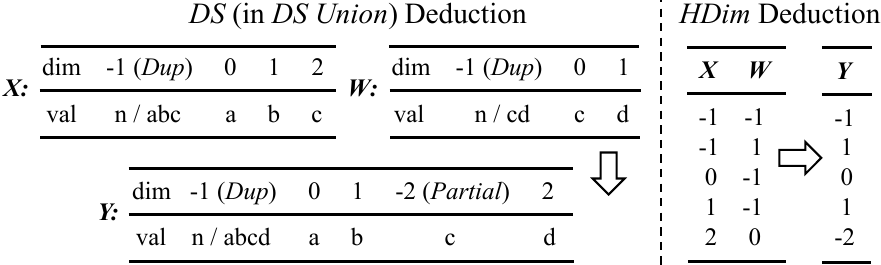}
\myvspace{-20pt}
\caption{\small{The deduction rules
of \DistributedStatesUnion and \HeteroDim for a 3D $\times$ 2D \texttt{Dot} operator. $n$ is the number of GPUs in the \textit{sharding subgroup} and $a$-$c$ are values along different sharding dimensions.}}
\label{fig:deduc}
\myvspace{-10pt}
\end{figure}

As shown in Figure~\ref{fig:graph_specialization}, we start by deducing the unspecified annotations (e.g., the output tensor $X$ of \texttt{Gelu} and the output tensor $Y$ of \texttt{Dot}). For simple unary operators like \texttt{Gelu}, the input tensor's annotation is straightforwardly propagated to the output. However, more complex operators, such as \texttt{Dot}, require a more intricate deduction process. Below, we outline the general methodology of annotation deduction.

\mysubsubsection{Deduction of \textit{DG Union} and \textit{HSize}}
For any standard operator, the output tensors inherit the same \DeviceGroupUnion and \HeteroSize as the inputs. This requirement stems from a semantic constraint enforced by \system: all inputs and outputs of an operator must have matching \DeviceGroupUnion and \HeteroSize, ensuring that the operator executes locally on its assigned device, consuming inputs to produce outputs without implicit cross-device resharding. For multi-input operators with mismatched \DeviceGroupUnion or \HeteroSize, users must explicitly insert \texttt{Reshard} operators before certain inputs to resolve the discrepancy.

\mysubsubsection{Deduction of \textit{DS Union} and \textit{HDim}}
The deduction of \DistributedStatesUnion and \HeteroDim depends on the specific characteristics of each operator and is governed by the rules of distributed computation. To illustrate this process, we analyze the \texttt{Dot} operator, which takes a 3D tensor $X$ and a 2D tensor $W$ as inputs, as a concrete example. In the \DistributedStatesUnion deduction process (Figure~\ref{fig:deduc}, left), since the \DeviceGroupUnion and \HeteroSize are aligned, the deduction simplifies to sequentially deriving the \DistributedStates for each \textit{sharding subgroup}. This mirrors the SPMD deduction, where sharding values for $X$, $W$, and $Y$ are derived using straightforward, rule-based logic. As for \HeteroDim deduction, top-tier sharding is essentially a simplified 1D sharding, following a similar rule-based logic. For example, as shown in Figure~\ref{fig:deduc} (right), if $X$ has a \HeteroDim of 0 and $W$ has a \HeteroDim of -1, the output tensor $Y$ retains a \HeteroDim of 0, preserving the heterogeneous top-tier \textit{Split} after the \texttt{Dot} operator.

\mysubsubsection{Discussions}
We support annotation deduction for a wide range of operators. As shown in Table~\ref{tb:annotation_deduction}, for the vast majority, the annotation is propagated from inputs to outputs without modification. Deviations from this default behavior are limited: only \texttt{Reshard} may alter \DeviceGroupUnion and \HeteroSize, while changes to \DistributedStatesUnion and \HeteroDim are confined to a few specific operators (e.g., \texttt{Dot}) that inherently transform sharding patterns and thus employ custom deduction logic. This design, where most operators reuse a common propagation procedure, allows \system's deduction to be efficient and scalable.

\myvspace{-5pt}
\subsection{Operator Instantiation}
\label{subsec:ops_inst}

Upon obtaining the annotations for all tensors, we instantiate a unique \textit{executable graph} for each device. 
This process primarily consists of two key steps.

\mysubsubsection{Non-Local operator removal}
First, operators whose inputs and outputs do not involve the local device in their \DeviceGroupUnion are excluded from its \textit{executable graph}. For instance in Figure~\ref{fig:graph_specialization}, since no tensors prior to $Y'$ are placed on GPU 6, all operators except \texttt{Reshard} (id=2) are removed. This pruning step is essential for enabling pipeline execution (\S\ref{subsec:exec}).

\begin{table}[!t]
\centering
\caption{\small{Operator annotation deduction behaviors.}}
\label{tb:annotation_deduction}
\myvspace{-8pt}
\small
\begin{tabular}{c|c}
\hline
\toprule
Operators & Modified Annotation Attributes \\
\midrule
\texttt{Reshard} & \DeviceGroupUnion, \HeteroSize, \DistributedStatesUnion, \HeteroDim  \\

\specialcell{\texttt{Dot}, \texttt{Sum}, \texttt{Reshape}, etc.} & \DistributedStatesUnion, \HeteroDim \\

\specialcell{All other operators} & — \\
\bottomrule
\hline
\end{tabular}
\myvspace{-10pt}
\end{table}

\mysubsubsection{Reshard instantiation} For all remaining \texttt{Reshard} operators, we apply the hierarchical communication resolution (\S\ref{sec:comm}) to instantiate them with the exact communication operators. 

\textbf{(\uppercase\expandafter{\romannumeral1}) Bottom-tier communication.} For bottom-tier communication, each \textit{sharding subgroup} independently determines its communication operator and instantiates it accordingly: In Figure~\ref{fig:graph_specialization}, \texttt{Reshard} (id=2) is classified as bottom-tier, leading GPU 0 and GPU 6 (belonging to different \textit{sharding subgroups}) to instantiate it with \texttt{RS} and \texttt{BSR}, respectively.

\textbf{(\uppercase\expandafter{\romannumeral2}) Top-tier communication.}  For operators classified as top-tier, all GPUs in the \DeviceGroupUnion apply the same instantiation logic: In Figure~\ref{fig:graph_specialization}, \texttt{Reshard} (id=1) is identified as \texttt{SplitAG} and uniformly instantiated on GPU 0–4.

\myvspace{-5pt}
\subsection{Execution}
\label{subsec:exec}

At runtime, operators in the \textit{executable graphs} are not invoked directly. Instead, they are organized into pre-/post-processing steps and forward/backward passes, which are repeatedly scheduled in a pipelined fashion. 
\system supports various scheduling strategies (e.g., GPipe~\cite{GPipe}, 1F1B~\cite{PipeDream}) and allows pipelines to independently process different micro-batch counts and sizes (Figure~\ref{fig:graph_specialization}, right). While the \textit{annotation plan} defines the sharding patterns and indicates the \texttt{Reshard} behaviors (Figure~\ref{fig:workflow}, line 1), the concrete shapes of the input data and the intermediate tensors (e.g., tensors to be communicated) are determined dynamically on each device at runtime (Figure~\ref{fig:workflow}, line 9), independent of the static \textit{executable graphs}. This decoupling between graph specialization and runtime execution enables flexible pipeline scheduling and supports variable-length inputs. Due to the space constraint, we refer interested readers to Appendix~\ref{appendix:exec} for more details.

\subsection{Discussions}
\label{subsec:graph_spec_discuss}
Below we summarize the data structures maintained per device (i.e., GPU). Following the SPMD paradigm, \system runs one process per device, and each process independently maintains its own copy of:
\textbf{(\lowercase\expandafter{\romannumeral1})} a device-agnostic \textit{defined graph} of the user program,
\textbf{(\lowercase\expandafter{\romannumeral2})} a device-agnostic \textit{annotation plan},
\textbf{(\lowercase\expandafter{\romannumeral3})} the corresponding device-specific \textit{executable graph} derived via graph specialization,
\textbf{(\lowercase\expandafter{\romannumeral4})} operator metadata, including the \textit{BSR tables} for all \texttt{BSR} operators in its \textit{executable graph}.
Among these, \textbf{(\lowercase\expandafter{\romannumeral1})} and \textbf{(\lowercase\expandafter{\romannumeral2})} are identical in content across processes, while \textbf{(\lowercase\expandafter{\romannumeral3})} and \textbf{(\lowercase\expandafter{\romannumeral4})} are unique to each process. 
Since these structures encode only parallelization and execution logic (excluding model parameters and activations), each process keeps them in the CPU DRAM of its host. In practice, they occupy only a few MiB per CPU host.

\myvspace{-5pt}
\section{Dynamic Graph Switching}
\label{sec:graph_switching}

\system's graph specialization abstraction addresses \textit{spatial} heterogeneity by generating device-specific parallel execution logic given an \textit{annotation plan}. However, \textit{temporal} heterogeneity, which evolves over time, demands dynamic reconfiguration of the overall parallelization. This necessitates switching from one \textit{annotation plan} to another (Figure~\ref{fig:workflow}(b),(c)).

\begin{figure}[!t]
\centering
\includegraphics[width=\linewidth]{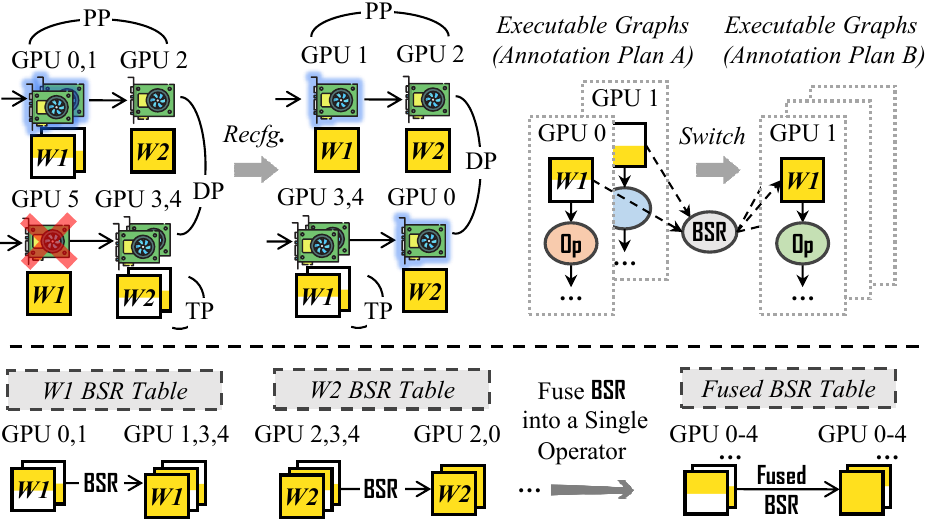}
\myvspace{-20pt}

\caption{\small{
Top: Online reconfiguration is achieved by transforming weight annotations using \texttt{BSR} across \textit{executable graphs}.
Bottom: Fusing \texttt{BSR} into a single operator further enhances efficiency.}}
\label{fig:graphs_switching}
\myvspace{-10pt}
\end{figure}

As illustrated in Figure~\ref{fig:graphs_switching} (top), to enable fast reconfiguration, we avoid reloading from checkpoints under the new sharding format. Instead, we reshard the weights from the old \textit{executable graphs} by leveraging the high interconnect bandwidth among devices. Since weight annotations do not involve \textit{Partial}, each weight can be directly resharded using the \texttt{BSR} operator (\S\ref{subsec:bsr}) to bridge the old and new plans.

To further optimize communication, rather than executing a separate \texttt{BSR} operator for each weight, we consolidate all \textit{BSR tables} into a fused one (Figure~\ref{fig:graphs_switching}, bottom), so that the corresponding \textit{fused BSR scheme} can balance the communication load across all GPUs. Moreover, we fuse multiple send-receive operators between the same GPU pair into a single operator, significantly reducing kernel launch overhead. By executing this \texttt{Fused BSR} operator, the system can therefore efficiently transition to the new \textit{annotation plan} and resume execution without restart or checkpoint reloading.\footnote{If a slice is missing from the fused \textit{BSR table} (i.e., no GPU currently owns that slice), reconfiguration falls back to checkpoint reloading.}

\mysubsubsection{Discussions}
A natural concern is whether maintaining the data structures of different parallel strategies (i.e., multiple plans and graphs) incurs prohibitive overhead as the system scales to large and heterogeneous clusters. The answer is no. A key observation is that the number of strategies is independent of cluster scale or heterogeneity.
For unstable devices, only a single strategy is maintained at any time, as the previous one is discarded after each reconfiguration. For mixed-length data, we pre-generate a set of strategies, each optimized for a specific maximum sequence length (e.g., one for batches with maximum length up to 16K and another up to 32K), and select the appropriate strategy at runtime based on the actual maximum sequence length of each batch. As a result, the number of strategies scales only with the granularity of the maximum-sequence-length partitioning, rather than with the cluster itself. Consequently, the overall maintenance overhead remains bounded regardless of cluster scale or heterogeneity.

\myvspace{-5pt}
\section{\system's Planner}
\label{sec:impl}

\begin{figure*}[!t]
\centering
\includegraphics[width=\textwidth]{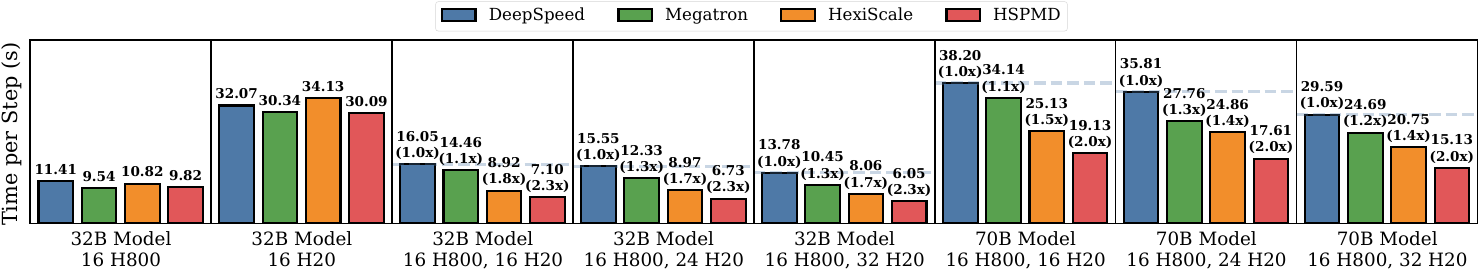}
\myvspace{-20pt}
\caption{\small{Scenario (a): Training performance across different device configurations and model size.}}
\label{fig:mfu}
\myvspace{-15pt}
\end{figure*}

This section presents the key implementation aspects of \system's planner, which automatically generates \textit{annotation plans} and thus frees users from constructing them by hand. In \system, the planner is treated as an external component that can be readily replaced by alternative implementations.

\mysubsubsection{Cost model and plan generation}
Following existing works on automatic parallelism~\cite{Alpa,Unity,Galvatron}, our default implementation adopts a profiling-based approach, building a \textit{cost model} for each operator on each hardware resource: for example, a compute operator (e.g., \texttt{Attention}) on a specific device (e.g., H20 or H800 GPU), or a communication operator (e.g., \texttt{AR}, \texttt{BSR}) on a given interconnect (e.g., NVLink or IB). 
We profile representative configurations (e.g., model sizes, tensor parallel degrees, input lengths) and fit cost models. To evaluate a plan, the planner decomposes its dataflow into operators and aggregates their predicted costs. Based on these estimates, it generates optimized \textit{annotation plans} using ILP, MINLP, or dynamic programming, implemented with PuLP~\cite{pulp} and Pyomo~\cite{pyomo}. Details are deferred to Appendix~\ref{appendix:plan}.

\mysubsubsection{Overhead of the planner}
An \textit{annotation plan} is a JSON dictionary with negligible storage cost. The planning consists of an offline phase (pre-generation) and an online phase (generation or selection). As shown in Figure~\ref{fig:workflow}, the offline phase runs before the training loop and is therefore off the critical path, while the online phase depends on the scenario:
\textbf{(a)} heterogeneous devices require no online planning,
\textbf{(b)} unstable devices trigger only a single re-generation upon cluster changes;
\textbf{(c)} mixed-length data uses pre-generated plans with lightweight online selection.
The overhead is therefore well controlled, with a detailed breakdown provided in Appendix~\ref{appendix:plan}.

\mysubsubsection{Generalizability to evolving models and clusters}
The \textit{cost model} is operator-centric rather than configuration-specific, so it generalizes naturally across model architectures and cluster settings. Model structural properties (e.g., layer count) and cluster factors (e.g., device count and topology) therefore do not affect it. Re-profiling is only required when new operators or new hardware resources are introduced (e.g., a new compute operator, a new GPU type, or a new interconnect).

\mysubsubsection{Sensitivity to annotation plan quality}
End-to-end performance is directly influenced by the quality of the \textit{annotation plan}.
Nevertheless, a clear lower bound is always guaranteed: in the extreme case, the planner can fall back to a fully homogeneous strategy, under which \system reduces to standard SPMD execution. Since our goal is to establish \system as a general and flexible framework for heterogeneous execution rather than to optimize scenario-specific planners, improving planner optimality is left to future work.

\myvspace{-5pt}
\section{Evaluations}
\label{sec:exp}

In this section, we evaluate \system under three representative scenarios: (a) heterogeneous devices, (b) unstable devices, and (c) mixed-length data (the workflow of each is depicted in \S\ref{sec:workflow}). Our evaluation aims to answer two central questions:
\begin{itemize}[noitemsep, topsep=0pt, parsep=0pt, partopsep=0pt, leftmargin=*]
\item How does \system outperform standard SPMD in the presence of diverse \textit{spatial} and \textit{temporal} heterogeneity?
\item Why does extending low-level \textit{declarative annotations}, as in \system, provide greater effectiveness than relying on high-level schedulers, as in other SPMD variants?
\end{itemize}

\mysubsubsection{Baselines}
As shown in Table~\ref{tab:baselines}, we select two categories of baselines. The first category includes standard SPMD training systems:
\textbf{(\lowercase\expandafter{\romannumeral1})} DeepSpeed~\cite{Zero, DeepSpeedSP}, which supports DP (with ZeRO-series) and SP;
\textbf{(\lowercase\expandafter{\romannumeral2})} Megatron~\cite{Megatron-LM, MegatronSP, UsingMegatron-LM}, which supports DP (with ZeRO-1), TP, PP, and CP.
The second category consists of scenario-specific systems that enhance SPMD with dedicated schedulers:
\textbf{(\lowercase\expandafter{\romannumeral3})} HexiScale~\cite{HexiScale}, a framework built on top of Megatron, employing a heterogeneous GPipe~\cite{GPipe} scheduler to support varying pipeline stages and TP degrees across stages;
\textbf{(\lowercase\expandafter{\romannumeral4})} Oobleck~\cite{Oobleck}, an elastic training framework that maintains fault tolerance with pre-defined pipeline templates, and features a scheduler for dynamic reconfiguration by merging and borrowing resources between pipeline templates;
\textbf{(\lowercase\expandafter{\romannumeral5})} HotSPa~\cite{HotSPa}, which utilizes a hot-switchable scheduler to reconfigure between several pre-defined homogeneous strategies (DP, TP, PP) to handle sequences of varying lengths, without incurring the overhead of cold-start.

\begin{table}[!t]
\centering
\caption{\small{Baselines: Two scenario-agnostic SPMD systems, and three scheduler-enhanced SPMD variants designed for specific scenarios.}}
\label{tab:baselines}
\myvspace{-8pt}
\small
\begin{tabular}{c|c|c}
\hline\toprule
Systems & Scenarios & Paradigms \\
\midrule
\specialcell{\textbf{(\lowercase\expandafter{\romannumeral1})} DeepSpeed\\\textbf{(\lowercase\expandafter{\romannumeral2})} Megatron} &
\specialcell{(a)-(c)}
& \specialcell{Standard SPMD} \\
\midrule
\textbf{(\lowercase\expandafter{\romannumeral3})} HexiScale & 
\specialcell{Only (a)}
& \specialcell{+ Hetero. Pipeline Scheduler} \\
\hline
\textbf{(\lowercase\expandafter{\romannumeral4})} Oobleck & \specialcell{Only (b)}
& \specialcell{+ Elastic Pipeline Scheduler} \\
\hline
\textbf{(\lowercase\expandafter{\romannumeral5})} HotSPa & 
\specialcell{Only (c)} & \specialcell{+ Hot-Switchable Scheduler} \\
\bottomrule\hline
\end{tabular}
\myvspace{-10pt}
\end{table}

\mysubsubsection{Experimental setup}
Our testbed consists of 16 H800 and 32 H20 GPUs (detailed in Appendix~\ref{appendix:cluster}). Given that LLMs represent the most significant DL workload today, and that existing baselines predominantly focus on LLMs, our evaluation centers on them as well. We adopt the widely‑used Llama architecture~\cite{Llama2} in various model sizes, with a default context length of 4K and a global batch size of 64, measuring per‑step training time. All baselines are carefully tuned to their optimal performance. For \system, we employ profiling‑based planners (detailed in Appendix~\ref{appendix:plan}) to generate and select the \textit{annotation plan}. Strategies are provided in Appendix~\ref{appendix:exp}.

\mysubsubsection{(a) Heterogeneous devices}
We first evaluate performance under static \textit{spatial} heterogeneity through experiments on heterogeneous devices. As shown in Figure~\ref{fig:mfu}, on homogeneous devices, all systems perform similarly, confirming that performance differences do not stem from engineering artifacts. On heterogeneous devices, however, \system consistently outperforms the baselines. DeepSpeed and Megatron are constrained by symmetric sharding, which hinders workload balance on devices with varying capabilities. HexiScale, on the other hand, has two key limitations: \textbf{(\lowercase\expandafter{\romannumeral1})} its built-in scheduler tightly couples the heterogeneous parallel strategy specification with execution, making it difficult to flexibly support complex pipeline scheduling schemes like 1F1B; and \textbf{(\lowercase\expandafter{\romannumeral2})} its scheduler lacks support for generalized asymmetric communication, relying instead on coarse-grained broadcasts. In contrast, \system leverages \textit{declarative annotations}, which decouple heterogeneous specifications from automatic pipeline execution (\S\ref{subsec:exec}) and enable hierarchical communication constructs (\S\ref{sec:comm}), thus facilitating flexible pipeline scheduling and efficient communication, resulting in improved performance.

\begin{figure}[!t]
\centering
\includegraphics[width=\linewidth]{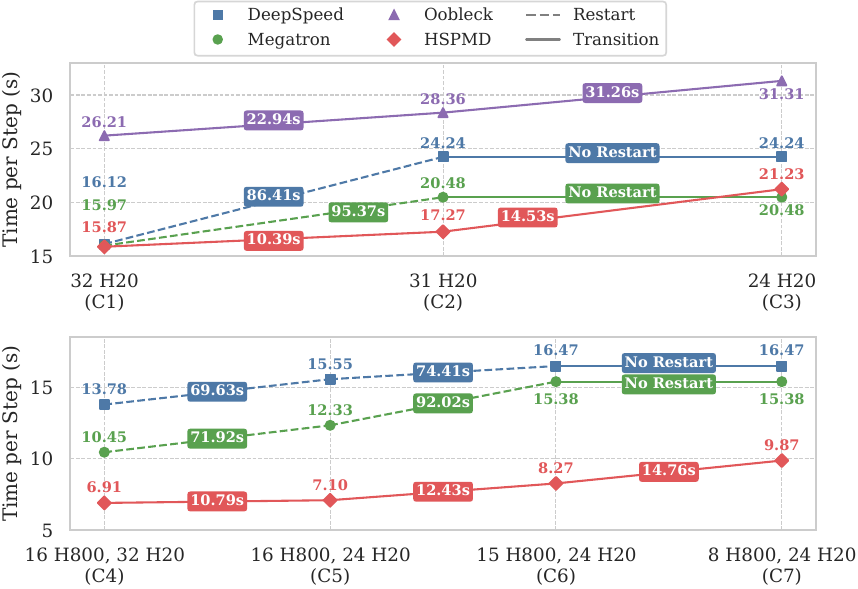}
\myvspace{-20pt}
\caption{\small{Scenario (b): Elastic training traces on homogeneous and heterogeneous clusters. We annotate the per-step training time for each configuration, along with the reconfiguration overhead.}}
\label{fig:elastic}
\myvspace{-10pt}
\end{figure}

\mysubsubsection{(b) Unstable devices}
We next evaluate elastic training performance under \textit{spatial} and unknown \textit{temporal} heterogeneity. Notably, DeepSpeed and Megatron employ a checkpoint-and-restart approach upon reconfiguration and use optimal training strategies under each configuration. Both \system and Oobleck~\cite{Oobleck} disable ZeRO-1~\cite{Zero} to retain DP-dimension redundancy for recovery without checkpoint-and-restart. This design is common across fault-tolerant training systems (including other recent works like Recycle~\cite{ReCycle}). Thus, both our generalized system and the specialized baseline inherently trade some training performance for fault tolerance.
As shown in Figure~\ref{fig:elastic}, we conduct experiments training a 32B model using two traces, each incorporating GPU and node failures.

Upon a GPU failure, DeepSpeed and Megatron are constrained by their symmetric sharding and must discard the entire node (\textit{spatial}). In contrast, \system can leverage all remaining GPUs, leading to improved performance (e.g., C2). Moreover, their significant restart overhead further makes them less suitable for elastic environments (\textit{temporal}).

On the other hand, Oobleck enables restart-free reconfiguration but underperforms compared to other systems because its elastic pipeline scheduler confines fault tolerance strategies to predefined pipeline templates. By comparison, \system employs tensor-level \textit{declarative annotations} rather than rigid pipeline-level templates to specify parallelization, which enables exploration of a much broader strategy space. Furthermore, while Oobleck’s elastic scheduler only supports na\"ive model broadcasting during reconfiguration, \system’s \textit{declarative annotations} allow more general per-tensor send-receive analysis to form a more balanced \texttt{Fused BSR} (as evaluated in \S\ref{sec:case_study}),  significantly reducing reconfiguration overhead.

\begin{figure}[!t]
\centering
\includegraphics[width=\linewidth]{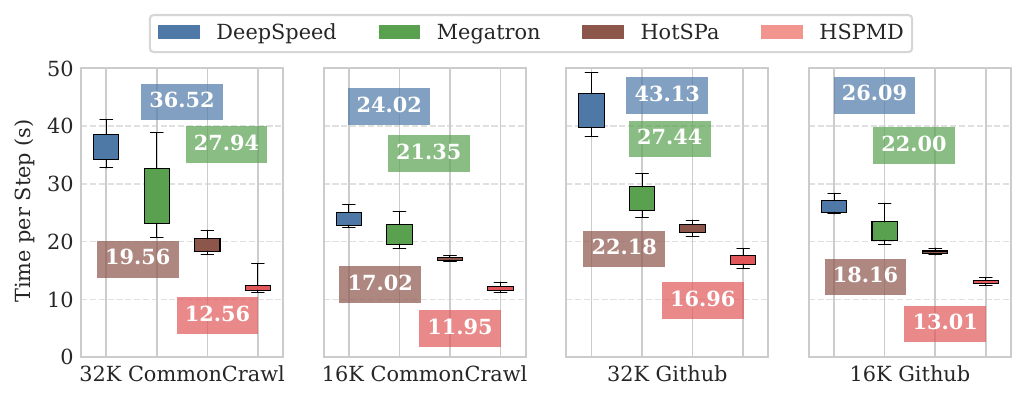}
\myvspace{-20pt}
\caption{\small{Scenario (c): Training performance with mixed-length data using different context lengths and datasets on 32 H20 GPUs. Box plots show time distributions with mean values annotated.}}
\label{fig:varlen_e2e}
\myvspace{-10pt}
\end{figure}

\begin{figure}[!t]
\centering
\includegraphics[width=\linewidth]{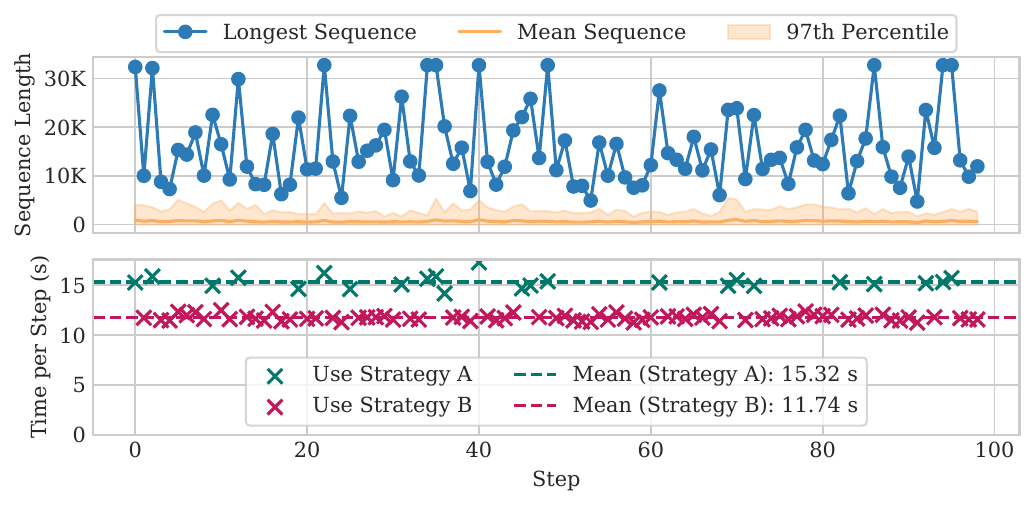}
\myvspace{-22pt}
\caption{\small{Sequence length variation and strategies employed by \system in the ``32K CommonCrawl'' case. \system dynamically switches between heterogeneous strategies (A and B) across different steps to optimize performance when sequence lengths shift.}}
\label{fig:seq_and_time}
\myvspace{-10pt}
\end{figure}

\mysubsubsection{(c) Mixed-length data} Finally, we evaluate the mixed-length data scenario to assess performance under \textit{spatial} and predictable \textit{temporal} heterogeneity. We train a 32B model for 100 steps on CommonCrawl and GitHub datasets using 32 H20 GPUs with a batch size of 200K tokens, testing different context lengths (32K and 16K). As baselines, DeepSpeed and Megatron pack mixed-length sequences into the fixed context window, truncating any excess~\cite{Packing}. HotSPa also employs data packing and additionally switches between pre-defined strategies for different length intervals. Results are shown in Figure \ref{fig:varlen_e2e}, while Figure \ref{fig:seq_and_time} illustrates a specific case: \system pre-generates two heterogeneous strategies—Strategy A, which is slower but handles longer sequences, and Strategy B, optimized for shorter ones. \system dynamically selects and switches between them across training steps (as depicted in Figure~\ref{fig:workflow}(c)). All reported per-step timings include strategy reconfiguration overhead for both HotSPa and \system.

DeepSpeed and Megatron, constrained by the symmetric nature of SPMD to a fixed homogeneous strategy, perform poorly because most sequences are much shorter than the context limit (97\% < 8K in Figure \ref{fig:seq_and_time}), making long-context-oriented strategies inefficient for packed short sequences. While HotSPa’s hot-switchable scheduler allows dynamic adjustment to length-tailored parallel strategies in the temporal dimension, it remains constrained by the symmetry of SPMD and cannot support spatially heterogeneous parallel strategies. This leads to lower performance compared to \system, which addresses both \textit{spatial} and \textit{temporal} heterogeneity through a unified abstraction of extended 
 \textit{declarative annotations}.

\mysubsubsection{Summary} Our evaluation across diverse scenarios reveals two findings: \textbf{(\lowercase\expandafter{\romannumeral1})} standard SPMD is inadequate for tackling heterogeneity, and \textbf{(\lowercase\expandafter{\romannumeral2})} scheduler-enhanced SPMD variants can address specific scenarios but have intrinsic limitations. First, compared to primitive-level extensions, addressing heterogeneity at the scheduler level is coarse-grained, which constrains the design space of heterogeneous strategies and the flexibility of strategy reconfiguration (e.g., HexiScale cannot support flexible pipeline schedulers, Oobleck relies on fixed pipeline templates, and HotSPa uses rigid homogeneous strategies), thereby limiting their performance. More importantly, these approaches work around SPMD’s symmetry by building a top-down system stack tailored to particular scenarios, making them effective only in isolated cases. In contrast, HSPMD introduces low-level extensions on \textit{declarative annotations}, directly bridging the gap between SPMD's symmetric nature and the asymmetric demands of heterogeneous environments. This enables a versatile, bottom-up system stack that provides general solutions across diverse scenarios.

\begin{figure}[!t]
\centering
\includegraphics[width=\linewidth]{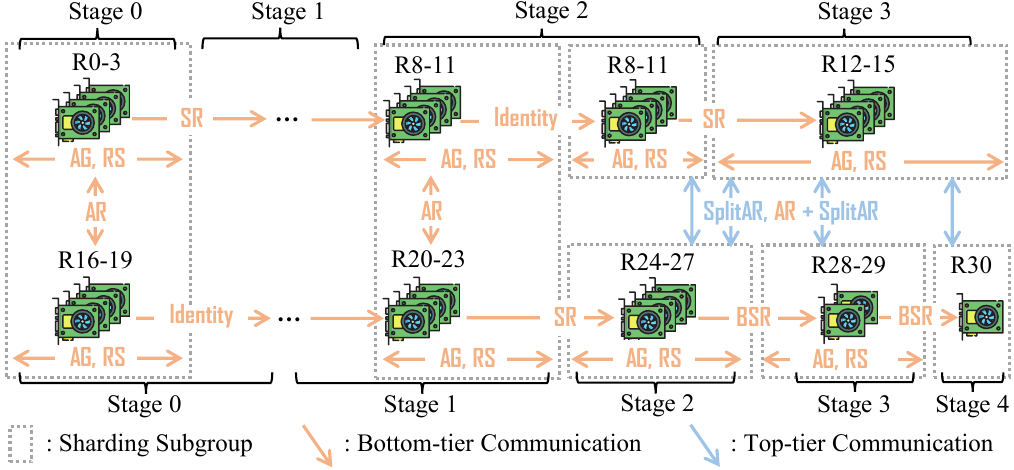}
\myvspace{-20pt}
\caption{\small{Parallel strategy employed in C2. ``R'' represents rank.}} 
\label{fig:case}
\myvspace{-10pt}
\end{figure}

\myvspace{-5pt}
\section{Case Study}
\label{sec:case_study}

In this section, we provide an in-depth analysis of HSPMD’s components, using the C2 configuration and the C1-to-C2 reconfiguration  process (in Figure~\ref{fig:elastic}) as key examples.

\mysubsubsection{Strategy deployment and communication resolution}
Figure~\ref{fig:case} shows \system’s deployment and communication patterns under C2 (31 H20 GPUs), forming two independent pipelines. Most stages use 4 GPUs, except the last two in the second pipeline, which use 2 and 1 GPU(s). Within each stage, TP is applied via \texttt{AG} and \texttt{RS} (\S\ref{subsec:bottom_comm}(\uppercase\expandafter{\romannumeral2})). And inter-stage communication within a pipeline employs \texttt{SR} (\S\ref{subsec:bottom_comm}(\uppercase\expandafter{\romannumeral1})) or \texttt{BSR} (\S\ref{subsec:bsr}). For cross-pipeline gradient synchronization, \texttt{AR} (\S\ref{subsec:bottom_comm}(\uppercase\expandafter{\romannumeral2})), \texttt{SplitAR} (\S\ref{subsec:top_comm}(\uppercase\expandafter{\romannumeral1})), and subgroup-specific \texttt{AR} following \texttt{SplitAR} (\S\ref{subsec:top_comm}(\uppercase\expandafter{\romannumeral2})) are used. Besides, we also provide the loss curves of C1 and C2 in Appendix~\ref{appendix:loss}, which further confirm that the heterogeneous strategy and the introduced communication operators do not affect convergence.

\begin{figure}[!t]
\centering
\includegraphics[width=\linewidth]{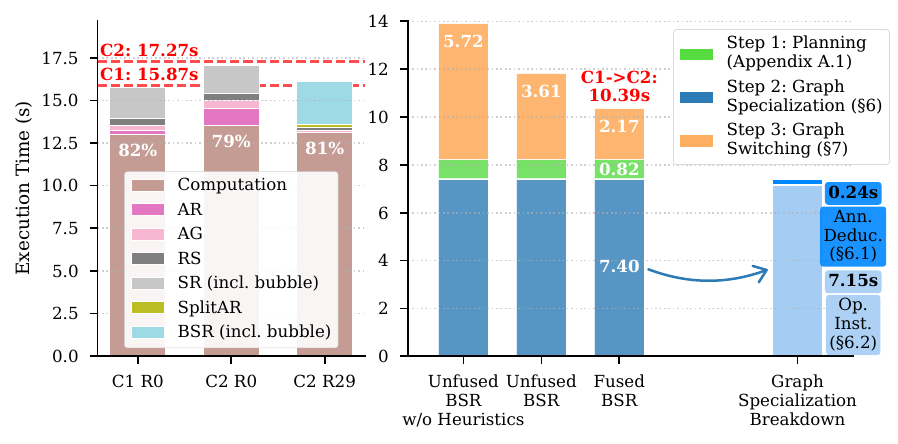}
\myvspace{-25pt}
\caption{\small{Left: Time per step breakdown for homogeneous (C1) and heterogeneous (C2) parallel strategies. Right: Reconfiguration overhead from C1 to C2 with different \texttt{BSR} algorithms.}}
\label{fig:breakdown}
\myvspace{-5pt}
\end{figure}

\mysubsubsection{Strategy execution time breakdown}
We then analyze the per-step computation and communication time for each rank in C2. Figure~\ref{fig:breakdown} (left) contrasts ranks 0 and 29, which follow asymmetric execution logic, against rank 0 in the homogeneous C1 as a reference. The results reveal that: \textbf{(\lowercase\expandafter{\romannumeral1})}  the heterogeneous strategy achieves balanced workload distribution across ranks; and \textbf{(\lowercase\expandafter{\romannumeral2})}  similar to the homogeneous strategy, computation remains the dominant runtime component. The additional overhead introduced by \texttt{SplitAR} and \texttt{BSR} is minimal, indicating that \system's asymmetric communication does not incur significant performance degradation.

\mysubsubsection{Strategy reconfiguration overhead}
We eventually evaluate the overhead of reconfiguring from C1 to C2, which consists of three sequential phases (Figure~\ref{fig:workflow}(b)): planning (Appendix~\ref{appendix:plan}), graph specialization (\S\ref{sec:graph_specialization}), and graph switching (\S\ref{sec:graph_switching}). As shown in Figure~\ref{fig:breakdown} (right), the planner generates the \textit{annotation plan} with minimal latency. Graph specialization then follows, dominated by operator instantiation (\S\ref{subsec:ops_inst}), which involves adjusting the graph and creating new CCL communication groups but typically completes within 10s, while the cost of annotation deduction (\S\ref{subsec:anno_deduc}) is negligible.

As for graph switching, we compare three \texttt{BSR} algorithms: \textbf{(\lowercase\expandafter{\romannumeral1})}  A baseline without heuristics (using minimal rank IDs for broadcasting); \textbf{(\lowercase\expandafter{\romannumeral2})}  Non-fused per-tensor \texttt{BSR} with heuristics; \textbf{(\lowercase\expandafter{\romannumeral3})}  Our \texttt{fused BSR}. Results show that our approach achieves the lowest end-to-end switching overhead despite maintaining the same total communication volume. As shown in Table~\ref{tab:comm_vol}, this improvement stems from evenly distributed traffic across ranks and better utilization of high-bandwidth NVLink.

\myvspace{-5pt}
\section{Discussions and Future Work}
\label{sec:discussions}

In this section, we discuss two additional aspects of \system: \textbf{(\lowercase\expandafter{\romannumeral1})} its support for Mixture-of-Experts (MoE) models under expert imbalance, and \textbf{(\lowercase\expandafter{\romannumeral2})} its interoperability with existing SPMD frameworks. Both represent promising directions enabled by \system's design, which we leave to future work.

\begin{table}[!t]
\centering
\small
\caption{\small{Distribution of C1-to-C2 communication volume under different \texttt{BSR} algorithms. We show the total communication volume each rank sends via NVLink and InfiniBand (IB).}}
\myvspace{-8pt}
\textbf{Format:} \textit{Sender Rank ID: NVLink Vol. (MB)} $\mid$ \textit{IB Vol. (MB)} 
\vspace{0.1cm}

\begin{minipage}[t]{0.5\textwidth}
\centering
\textbf{Unfused BSR w/o Heuristics} 
\vspace{0.1cm} 
\begin{tabular}{*{4}{c}}
\hline
R4: 0$\,|\,$748 & R5: 0$\,|\,$747 & R6: 0$\,|\,$747 & R7: 0$\,|\,$747 \\
R8: 0$\,|\,$1495 & R9: 0$\,|\,$1495 & R10: 0$\,|\,$1495 & R11: 0$\,|\,$1495 \\ 
R12: 0$\,|\,$5517 & R13: 0$\,|\,$11496 & R14: 0$\,|\,$8222 & R15: 0$\,|\,$11496 \\ \hline
\end{tabular}
\end{minipage}
\quad
\begin{minipage}[t]{0.5\textwidth}
\centering
\textbf{Fused BSR} 
\begin{tabular}{*{4}{c}}
\hline
R8: 0$\,|\,$747 & R9: 0$\,|\,$747 & R10: 0$\,|\,$747 & R11: 0$\,|\,$747 \\ 
R15: 0$\,|\,$11496 & R20: 748$\,|\,$0 & R21: 748$\,|\,$0 & R22: 748$\,|\,$0 \\ 
R23: 748$\,|\,$0 & R24: 0$\,|\,$748 & R25: 0$\,|\,$748 & R26: 0$\,|\,$748 \\ 
R27: 0$\,|\,$748 & R28: 5517$\,|\,$0 & R29: 11496$\,|\,$0 & R30: 8222$\,|\,$0 \\ \hline
\end{tabular}
\end{minipage}
\label{tab:comm_vol}
\myvspace{-15pt}
\end{table}

\mysubsubsection{Support for MoE and expert imbalance}
MoE models are typically trained with expert parallelism (EP)~\cite{GShard}, which shards experts across devices, but they often suffer from load imbalance due to skewed token routing. A representative mitigation is EPLB~\cite{EPLB}, which replicates heavily loaded experts to distribute the workload. \system naturally supports such strategies through its two-tier sharding annotations: at the top tier, \system applies \textit{Split} on the hidden-size dimension to split across different experts, and at the bottom tier, it uses different \textit{Dup} degrees to replicate different loaded experts. When expert workloads shift substantially during training, \system can reconfigure the overall \textit{annotation plan} via graph switching, without requiring any further extension.

\mysubsubsection{Interoperability with existing SPMD frameworks}
Any prior annotation-based SPMD framework (e.g., Alpa~\cite{Alpa}, GSPMD~\cite{GSPMD}, and DTensor~\cite{PyTorchDTensor, TensorFlowDTensor}) can be extended to support \system's annotations by adding one additional tier on top of its existing SPMD annotations, which amounts to introducing two extra attributes (\HeteroDim and \HeteroSize) and promoting the original annotation structure to a list of such structures. On top of this, \system's communication resolution, graph specialization, and graph switching mechanisms can be plugged in to enable heterogeneous parallelization, while the framework's existing parallelization primitives (e.g., operator-level sharding rules, automatic annotation deduction, and its standard parallelism implementations such as DP, TP, PP, and CP) can be reused as-is. By contrast, frameworks that realize parallelism by specifying particular layers, such as Megatron~\cite{Megatron-LM} and DeepSpeed~\cite{Zero}, are harder to build \system on top of, since they do not expose a unified, operator-level annotation surface and instead rely on ad-hoc, per-layer specifications (e.g., \texttt{ColumnParallelLinear}).

\myvspace{-5pt}
\section{Related Work}
\label{sec:related_work}

\subsection{Tackling Workload Heterogeneity}

Recent works have tackled workload heterogeneity in large-scale DL training through scenario-specific optimizations.

\mysubsubsection{(a) Heterogeneous devices} HAP~\cite{HAP} uses a sharding ratio to handle uneven TP partitioning. Whale~\cite{Whale} partitions
workloads by splitting the global graph into heterogeneous
task graphs. AMP~\cite{AMP} and HetHub~\cite{HETHUB} support heterogeneous layer partitioning in PP, while Metis~\cite{Metis} extends Alpa~\cite{Alpa} for heterogeneous strategy search. HexiScale~\cite{HexiScale} enables more fine-grained heterogeneous 3D parallelism, and Sailor~\cite{Sailor} further optimizes the search cost for such strategies.

\mysubsubsection{(b) Unstable devices} Varuna~\cite{Varuna} uses checkpoint restarts and adjusts DP and PP to find optimal strategies after failures. Bamboo~\cite{Bamboo} achieves fault tolerance via redundant storage on adjacent pipeline stages, Oobleck~\cite{Oobleck} employs multiple pipeline templates for fault tolerance and strategy tuning, and ReCycle~\cite{ReCycle} rebalances the workload from failed pipelines.

\mysubsubsection{(c) Mixed-length data} FlexSP~\cite{FlexSP} uses dynamic SP for varied sequences, while ByteScale~\cite{ByteScale} and DCP~\cite{DCP} adopt dynamic CP. HotSPa~\cite{HotSPa} assigns homogeneous strategies to different length intervals.  WLB-LLM ~\cite{WLB-LLM} addresses load imbalance specifically in CP and PP, while Zeppelin ~\cite{Zeppelin} achieves finer-grained control over attention and NIC imbalance.

In contrast to these specialized solutions, \system introduces a unified abstraction via low-level \textit{declarative annotations} instead of high-level schedulers. This design offers greater generality and could serve as a foundational substrate for addressing diverse heterogeneity. Furthermore, prior works have proposed planning
algorithms tailored to targeted scenarios, which can be seamlessly integrated with our work by encoding their strategies through our annotations.

\myvspace{-5pt}
\subsection{SPMD and MPMD Training Systems}
Beyond the SPMD training systems discussed above, another line of work adopts the MPMD paradigm, such as Pathways~\cite{pathways}, JaxPP~\cite{JaxPP}, and Piper~\cite{Piper}. One motivation behind their MPMD design is the observation that pipeline parallelism (PP) is not inherently SPMD, as different devices naturally execute different model stages. Nevertheless, SPMD remains the dominant paradigm in mainstream training systems for its scalability and simplicity (see \S\ref{subsec:motivation}), and PP is typically encoded in an SPMD style, where devices share a global program but diverge at runtime. For example, GSPMD~\cite{GSPMD} treats PP as layer-wise sharding, while Megatron~\cite{Megatron-LM} employs a light scheduler to execute subgraphs. HexiScale~\cite{HexiScale} and Oobleck~\cite{Oobleck} further introduce more advanced schedulers for heterogeneous execution. These approaches effectively encode PP within SPMD. \system takes the same stance: rather than switching to MPMD, it extends SPMD with \textit{declarative annotations} for asymmetry, preserving single-program simplicity while enabling the heterogeneous execution of various parallelism schemes including PP (see \S\ref{subsec:exec}).

\myvspace{-5pt}
\section{Conclusion}
\label{sec:conclusion}

We present \system, a novel system that extends the SPMD paradigm to support heterogeneous parallelization and dynamic reconfiguration, addressing both \textit{spatial} and \textit{temporal} heterogeneity. Through low-level sharding annotations and hierarchical communication, \system bridges the gap between SPMD's inherent symmetric constraints and the emerging demands of asymmetric execution in large-scale DL training.
Extensive evaluation across (a) heterogeneous devices, (b) unstable devices, and (c) mixed-length data demonstrates \system's effectiveness and broad applicability.

\section*{Acknowledgments}
This work is supported by National Natural Science Foundation of China (U23B2048, 62402011), Fundamental and Interdisciplinary Disciplines Breakthrough Plan of the Ministry of Education of China (JYB2025XDXM108), and High-performance Computing Platform of Peking University. Fangcheng Fu and Bin Cui are the corresponding authors.

\bibliographystyle{plain}
\bibliography{reference}

\clearpage
\appendix
\section{Implementation Details}
\label{appendix:impl}

\begin{table*}[!t]
\centering
\small
\caption{\small{Workflow and time breakdown of different scenarios. Step~1 (planning) is automated by our planner module (users may optionally provide the \textit{annotation plan} manually), while Step~2 (graph specialization) and Step~3 (graph switching) are handled by \system. 
The planner generates the \textit{annotation plan} (offline/online) and may select the optimal plan if multiple candidates exist (online). \textit{Executable graphs} are specialized upon receiving the \textit{annotation plan} (offline/online), and different \textit{executable graphs} can be switched on-the-fly (online). Note that all online overheads are included in \S\ref{sec:exp}'s end-to-end results.}}
\myvspace{-5pt}
\begin{tabular}{P{3.5cm}P{4cm}P{4.5cm}P{4cm}}
\hline\toprule
\textbf{Scenario} & 
\makecell{\textbf{Step 1: Planning}} & 
\makecell{\textbf{Step 2: Graph} \\ \textbf{Specialization (\S\ref{sec:graph_specialization})}} & 
\makecell{\textbf{Step 3: Graph} \\ \textbf{Switching (\S\ref{sec:graph_switching})}} \\
\midrule
Heterogeneous Devices & 
\makecell{\textbf{Offline:} \\ Single \textit{annotation plan} \\ generated before training \\ (6.4s-18.2s).} & 
\makecell{\textbf{Offline:} \\ Executed once \\ before training \\ (6.8s-8.3s).} & 
\makecell{—} \\
\midrule
Unstable Devices & 
\makecell{\textbf{Offline:} \\ 
Initial \textit{annotation plan} \\ generated before training \\ (9.1s-18.2s). \\ \textbf{Online:} \\ 
New \textit{annotation plan} \\ generated whenever device \\ availability changes \\ (0.05s-1.2s).} & 
\makecell{\textbf{Online:} \\ Executed each time \\ a new \textit{annotation plan} \\ is generated \\ (6.2s-7.9s).} & 
\makecell{\textbf{Online:} \\ Triggered after \\ new \textit{executable graphs} \\ are specialized \\ (2.2s-5.3s, switch both \\ FP32 parameters \\ \& optimizer states).} \\
\midrule
Mixed-length Data & 
\makecell{\textbf{Offline:} \\ 
Multiple \textit{annotation plans} \\ generated before training \\ (24.1s-26.8s). \\ \textbf{Online:} \\ 
\textit{Annotation plan} selection \\ for next training steps \\ (1.5s-2.1s, overlapped with \\ the current step's training).} & 
\makecell{\textbf{Offline:} \\ Executed for \\ each \textit{annotation plan} \\ before training \\ (14.8s-16.2s).} & 
\makecell{\textbf{Online:} \\ Triggered between \\ training steps if \\ different \textit{annotation plans} \\ are favored \\ (0.8s-1.3s, switch only \\ BF16 parameters).} \\
\bottomrule\hline
\end{tabular}
\label{tab:plan}
\myvspace{-10pt}
\end{table*}

\system is designed to accommodate diverse forms of heterogeneity, which requires additional computation graph representations and dynamic adjustments. However, extracting computation graph information in PyTorch~\cite{Pytorch} is non-trivial. To address this, we have developed a prototype framework comprising 87.9K lines of C++ code and 16.9K lines of CUDA code, supporting graph-based distributed DL training. The framework implements all major parallelism paradigms, including data parallelism (DP), tensor parallelism (TP), pipeline parallelism (PP), and context parallelism (CP), as well as complementary techniques such as ZeRO-1, offloading, activation recomputation, and mixed-precision training.

Within the total 87.9K lines of C++, 16.9K lines implement the core logic of \system, covering sharding annotation, communication resolution, graph specialization, and graph switching. An additional 14.7K lines of C++, largely consisting of pybind~\cite{Pybind} bindings, serve as glue code that exposes \system APIs to Python. Together with 12.5K lines of Python, these APIs allow users to define tensors, operators, and training workflows, enabling seamless deployment and reconfiguration of heterogeneous strategies.

Our prototype system is optimized for large language models (LLMs) training, with collective communication primitives implemented via NCCL~\cite{NCCL} and computation kernels accelerated using libraries such as FlashAttention~\cite{FlashAttention,FlashAttention-2}, cuBLAS~\cite{cuBLAS}, and cutlass~\cite{cutlass}. When evaluated with homogeneous parallel strategies, \system achieves performance comparable to Megatron~\cite{Megatron-LM,MegatronSP,UsingMegatron-LM} (as evaluated in \S\ref{sec:exp}(a)). While our current implementation and experiments focus on LLMs—due to their extreme model sizes and demand for massive GPU clusters—the proposed designs are broadly applicable to other deep learning workloads.

Below, we provide additional details on the implementation that are not covered in the main text.

\subsection{Scenario-specific Planning}
\label{appendix:plan}

In this section, we present our approach for automatically generating and selecting \textit{annotation plans} (\S\ref{sec:workflow}) for the three representative scenarios in \S\ref{sec:exp}. The primary focus of this paper is on the underlying system design rather than on optimizing the planning process. To that end, the planner is implemented as a modular component that users can readily replace or extend—for example, by incorporating algorithms from prior scenario-specific work. The implementation described here represents just one of many possible planning algorithms, provided to demonstrate \system’s flexibility in supporting diverse application settings.

Our approach starts by profiling execution time and memory usage for homogeneous parallelism configurations on a single model unit (e.g., an \texttt{Attention} operator), given the target model and workload specifications (i.e., context length and batch size). Based on these measurements, we develop a regression-based \textit{cost model} to estimate per-device execution time and memory usage for any heterogeneous parallel strategy expressible in \system. This is achieved by analyzing the dataflow and aggregating results from different \textit{sharding subgroups}, each of which employs homogeneous parallelism and can be estimated individually. Below, we detail the algorithms used in each scenario. 

\mysubsubsection{(a) Heterogeneous devices}
Using our \textit{cost model}, we formulate a two-level optimization problem. The objective is to minimize the maximum estimated execution time across all devices while satisfying memory constraints:
\begin{itemize}[leftmargin=*]
    \item \textbf{Pipeline configuration} (first-level): We determine the number of pipelines (\S\ref{subsec:exec}) along with their parallel methods, stage counts, and device-to-stage mappings. This is formulated as a Mixed-Integer Nonlinear Programming (MINLP) problem.
    \item \textbf{Layer \& micro-batch assignment} (second-level): We assign model layers to different pipeline stages and calculate optimal micro-batch configurations (i.e., micro-batch size and count) for each pipeline, solved as multiple Integer Linear Programming (ILP) problems.
\end{itemize}
The combined solution yields a unique \textit{annotation plan} for deployment.

While the ILPs can be solved efficiently, the MINLP remains computationally expensive. Instead of solving the original MINLP directly, we adopt a heuristic-guided greedy search to prune suboptimal solutions upfront, thereby shrinking the strategy space and reducing the solving overhead while preserving solution quality.

\mysubsubsection{(b) Unstable devices}
Our elastic training approach begins by generating an optimal initial \textit{annotation plan} using the same planner as in (a). When reconfiguration is required, profiling is unnecessary since the model and workload remain unchanged. Moreover, we intelligently prune the MINLP search space using intermediate results from the initial planning phase. This optimization enables the first-level problem to be solved significantly faster during reconfigurations, while the second-level ILPs (remaining identical to the initial solution) continue to be resolved with minimal overhead.

\mysubsubsection{(c) Mixed-length data}
For datasets with variable sequence lengths, our methodology begins by analyzing the dataset distribution to obtain sequence length statistics. Based on these statistics, we employ a dynamic programming algorithm that jointly determines both the optimal sequence length intervals for partitioning variable-length sequences and the corresponding parallel method for each interval. Different pipelines, each tailored to a specific interval and parallel method, are then combined into a single heterogeneous parallel strategy. Recognizing that per-batch sequence length distributions may deviate from global dataset statistics, we prepare multiple parallel strategies optimized for different maximum sequence lengths. During execution, for each incoming batch, our system leverages the \textit{cost model} again to distribute sequences among pipelines and dynamically select the strategy with the lowest estimated execution time. This data assignment and strategy selection process runs concurrently on CPU resources, carefully orchestrated to overlap with GPU training. By parallelizing the planning of upcoming steps (e.g., preparing the next 10 steps during the current step’s execution) across available CPU cores, our implementation effectively amortizes and hides the planner’s runtime overhead.

Across all scenarios, we leverage PuLP~\cite{pulp} for solving ILPs and Pyomo~\cite{pyomo} for solving MINLPs. As detailed in Table~\ref{tab:plan}, we provide overhead measurements including planning, graph specialization (\S\ref{sec:graph_specialization}), and graph switching (\S\ref{sec:graph_switching}) for each scenario, accompanied by thorough workflow descriptions. The results indicate that our planners either introduce minimal overhead or run offline, ensuring negligible impact on online training.

\begin{figure*}[htbp]
\centering
\includegraphics[width=\linewidth]
{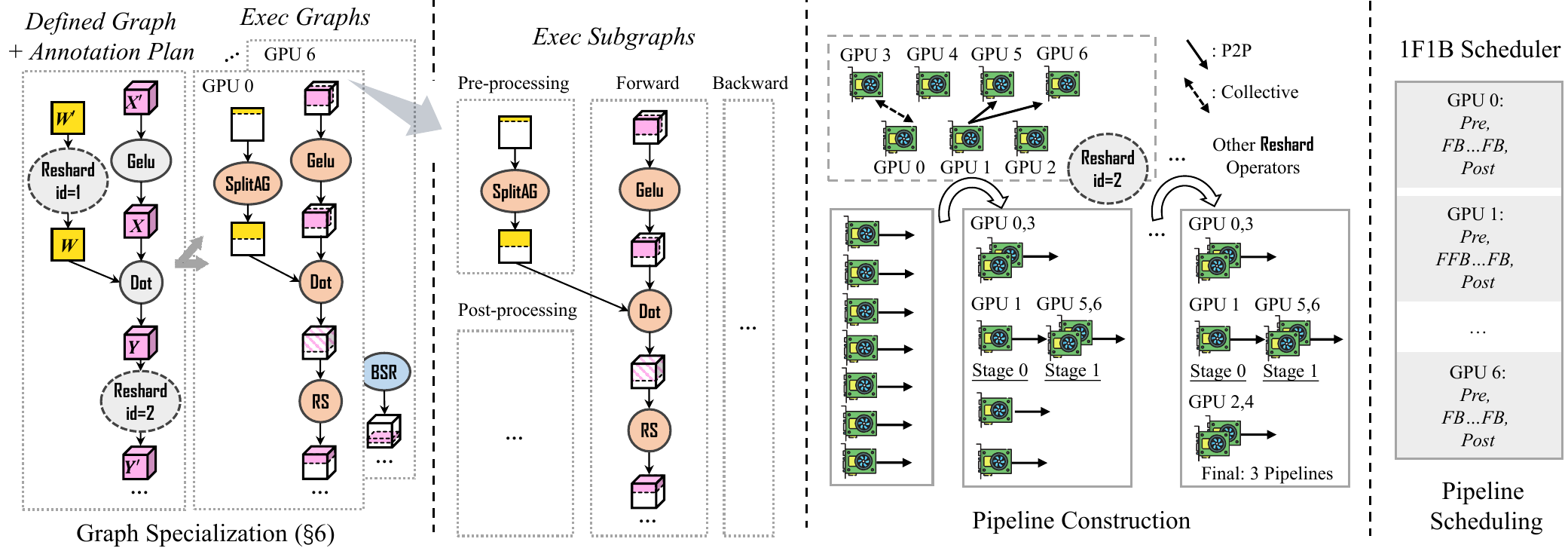}
\vspace{-15pt}
\caption{\small{Pipeline construction and scheduling. The \textit{executable graph} associated with each device is divided into multiple \textit{executable subgraphs}. Once the pipelines are constructed, each device is assigned a schedule that dictates the execution order of its \textit{executable subgraphs}. This figure aligns with Figure~\ref{fig:graph_specialization}.}}
\label{fig:pipe}
\end{figure*}

\subsection{Execution}
\label{appendix:exec}

In this section, we provide the details not covered in §\ref{subsec:exec} on how to recognize heterogeneous pipelines and coordinate their execution, as well as how to support dynamic input shapes during execution without relying on the sharding annotation system.

\mysubsubsection{Pipeline construction and scheduling}
\system supports various scheduling schemes, including GPipe~\cite{GPipe} and 1F1B~\cite{PipeDream}, while enabling independent pipelines to process micro-batches with different sizes and quantities. This functionality is achieved through two key steps: \textbf{(\lowercase\expandafter{\romannumeral1})} determining the pipeline structure based on current sharing annotations (i.e., identifying device-to-stage and device-to-pipeline mappings); and \textbf{(\lowercase\expandafter{\romannumeral2})} coordinating device execution according to the selected pipeline scheduler (e.g., 1F1B).

As illustrated in Figure~\ref{fig:pipe} (left), after \textit{executable graphs} are specialized (\S\ref{sec:graph_specialization}), we further partition each device’s \textit{executable graph} into four types of \textit{executable subgraphs}:

\begin{itemize}[leftmargin=*]
\item \textbf{Pre-processing} ($Pre$): Includes operators such as precision conversion (e.g., FP32 to BF16) for mixed-precision training, and parameter gathering (e.g., \texttt{AG}, \texttt{SplitAG}) for ZeRO~\cite{Zero} sharding.
\item \textbf{Post-processing} ($Post$): Includes operators for gradient accumulation (e.g., \texttt{Sum}), gradient synchronization (e.g., \texttt{RS}, \texttt{SplitRS}), and optimizer updates (e.g., Adam~\cite{Adam}).
\item \textbf{Forward} ($F$): Includes operators for the DL forward pass (e.g., \texttt{Dot}, \texttt{AttentionFwd}).
\item \textbf{Backward} ($B$): Includes operators for the DL backward pass (e.g., \texttt{Dot}, \texttt{AttentionBwd}).
\end{itemize}

These \textit{executable subgraphs} serve as the scheduling units for subsequent execution given the pipeline scheduler.

Below, we describe the process of constructing pipelines, whereby each device identifies its stage number and pipeline number, ensuring that all devices are aware of their roles during pipeline scheduling. We design an algorithm for constructing pipelines in a step-by-step manner. As illustrated in Figure~\ref{fig:pipe} (middle), we initially construct a separate pipeline for each device. Subsequently, we sequentially analyze all \texttt{Reshard} operators involved in the forward and backward \textit{executable subgraphs}. In Figure~\ref{fig:pipe}, we identify that \texttt{Reshard} (id=1) is executed only once at the beginning, which belongs to the pre-processing \textit{executable subgraph} and does not participate in forward or backward passes, so the first communication operator actually analyzed is \texttt{Reshard} (id=2). For this communication operator, we decompose its internal communication into two categories: collective communication and peer-to-peer (P2P) communication. All devices involved in collective communication are merged into the same pipeline, while devices involved in P2P communication are concatenated into subsequent stages of the pipeline. For example, in Figure~\ref{fig:pipe} (middle), after processing \texttt{Reshard} (id=2), the pipelines containing GPU 0 and GPU 3 are merged into one, while GPU 5 and GPU 6 are appended to the pipeline containing GPU 1 as the next stage. Through iterative application of this process, all pipelines are eventually fully constructed.

After pipeline construction, the system orchestrates execution according to the pipeline structure during runtime. Each device is assigned a schedule that defines the execution order of its \textit{executable subgraphs}. For instance, in Figure~\ref{fig:pipe} (right), when using the 1F1B scheduler, GPU 0 (located at the final stage) follows the schedule $Pre,FB \ldots FB,Post$, whereas GPU 1 (on the first stage) executes according to $Pre,FFB \ldots FB,Post$. Here, $Pre$ and $Post$ denote the pre- and post-processing \textit{executable subgraphs}, respectively, and $F$ and $B$ represent forward and backward \textit{executable subgraphs}, respectively.

\mysubsubsection{Symbolic shape extension}
During execution, \system dynamically adapts to variations in input tensor shapes across different pipelines (e.g., to facilitate load balancing, data packing, and other runtime optimizations). While static sharding annotations define the high-level sharding pattern, the concrete shapes of individual shards are resolved at runtime.

To support this dynamic behavior, \system extends tensor metadata with symbolic variables, which represent unknown dimensions of the shape (e.g., $\mathbf{S}$ for sequence length). These variables are propagated through tensor operations on each device while maintaining dimensional constraints. For instance, splitting a tensor along its sequence dimension would generate the constraint $\mathbf{S'} = \mathbf{S}/2$, preserving algebraic relationships between the symbols of the two tensors. The system resolves these symbols to concrete arithmetic values only when actual input tensors are provided at runtime to each device. This design allows different ranks to process dynamically shaped data. Additionally, \system enforces both compile-time and runtime checks to detect invalid symbol usage. For example, defining $\mathbf{S'} = \mathbf{S}/2$ without satisfying the constraint $\mathbf{S} \equiv 0 \ (\mathrm{mod}\ 2)$ is illegal and may lead to shape mismatches during cross-rank communication.

\section{Training Convergence}
\label{appendix:loss}

\begin{figure*}[htbp]
\centering
\includegraphics[width=\linewidth]{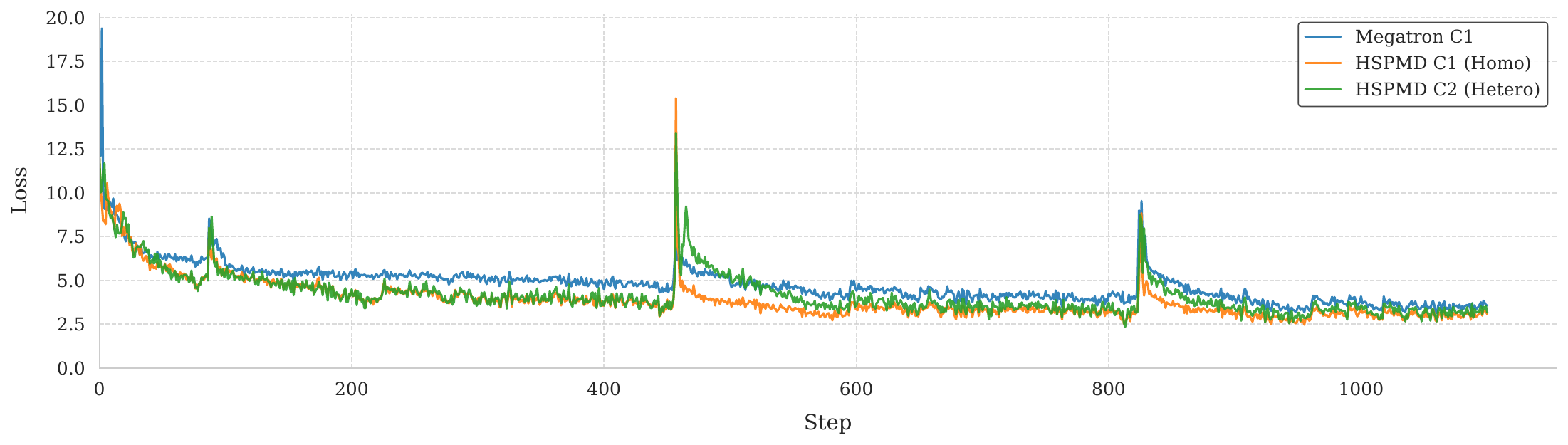}
\vspace{-20pt}
\caption{\small{Training loss comparison.}}
\label{fig:loss}
\end{figure*}

In this section, we present a comparative analysis of training loss under two configurations (C1 and C2, totally aligned with \S\ref{sec:case_study}). In both configurations, a 32B Llama model is trained using \system, yet they differ in GPU resources and parallel strategies:

\begin{itemize}[leftmargin=*]
    \item C1: The model is trained on 32 H20 GPUs under a homogeneous parallel strategy.
    \item C2: The model is trained on 31 H20 GPUs under a heterogeneous parallel strategy.
\end{itemize}
Comprehensive descriptions of these parallel strategies are documented in Appendix~\ref{appendix:elastic}, while the full training setup is summarized in Table \ref{tab:loss}.

\begin{table}[H]
\centering
\small
\caption{\small{Training setup.}}
\label{tab:llama32b_config}
\begin{tabular}{>{\bfseries}l>{\bfseries}lr}
\hline\toprule
\multicolumn{1}{l}{\textbf{Category}} & \multicolumn{1}{l}{\textbf{Parameter}} & \multicolumn{1}{r}{\textbf{Value}} \\
\midrule

\multirow{10}{*}{\textbf{Model Configuration}} 
& Number of Layers & 60 \\
& Hidden Size & 6,656 \\
& Attention Heads & 64 \\
& FFN Hidden Size & 17,920 \\
& Vocabulary Size & 50,304 \\
& Query Groups & 64 \\
& Activation Function & SwiGLU \\
& Normalization & RMSNorm \\
& Position Encoding & RoPE \\
& Dropout & None \\

\midrule

\multirow{14}{*}{\textbf{ Hyperparameters}} 
& Precision & BF16 \\
& Optimizer & Adam \\
& Learning Rate & $1 \times 10^{-4}$  \\
& Warm-up Steps & 0 \\
& Scheduler & Constant \\
& Weight Decay & 0 \\
& Adam $\beta_1$ & 0.9 \\
& Adam $\beta_2$ & 0.999 \\
& Adam $\epsilon$ & $1 \times 10^{-8}$ \\
& Gradient Clipping & None \\
& Dataset & Github \\
& Epoch & 1 \\
& Context Length & 4,096 \\
& Batch Size & 64 \\

\bottomrule\hline
\end{tabular}
\label{tab:loss}
\end{table}

We also train a model using Megatron, employing the same homogeneous parallel strategy and training setup as in C1. As shown in Figure~\ref{fig:loss}, the results show that \system achieves good convergence under both homogeneous and heterogeneous settings. In the homogeneous setting (C1), the convergence behavior closely matches that of the Megatron baseline, with minor differences in absolute loss values attributable to framework-level implementation variations. Besides, the heterogeneous setting (C2) exhibits convergence nearly identical to C1, with only marginal fluctuations observed during training irregularities. These findings demonstrate that the introduction of asymmetric sharding and diverse asymmetric communication patterns does not compromise training convergence.

\section{More Experimental Details}
\label{appendix:exp}

This section provides a comprehensive analysis of \system's heterogeneous parallel strategies across diverse scenarios and compares them against baseline methods. While these strategies (formulated as \textit{annotation plans}) are automatically generated by scenario-specific planners (Appendix~\ref{appendix:plan}), users also have the flexibility to define customized strategies by manually supplying their own \textit{annotation plan}.

\subsection{Cluster Setup}
\label{appendix:cluster}

Table~\ref{tab:gpu_config} presents the configuration of our computing cluster. The H800 GPU offers higher computational power but lower intra-node (NVLink) communication bandwidth, whereas the H20 GPU provides lower computational performance but higher communication bandwidth. Additionally, the two GPUs differ slightly in memory capacity, posing challenges for heterogeneous parallel computing on this setup.

\begin{table}[!t]
\centering
\small
\caption{\small{Heterogeneous GPU cluster.}}
\label{tab:gpu_config}
\begin{tabular}{lccc}
\hline\toprule
\textbf{GPUs} & \textbf{Memory} & \textbf{BF16 Tensor Core} & \textbf{NVLink} \\
\midrule
16 H800 & 81559 MB & 990 TFLOPS & 400 GB/s \\
32 H20 & 97871 MB & 148 TFLOPS & 900 GB/s \\
\bottomrule\hline
\end{tabular}
\end{table}

\subsection{Scenario (a): Heterogeneous Devices}
\label{appendix:hetero_cluster}

Table~\ref{tab:hetero_cluster} summarizes the optimal parallel strategies for DeepSpeed and Megatron, the two baselines, across different model sizes and heterogeneous device setups. These strategies were derived through an exhaustive search process, systematically evaluating various combinations of parallelisms and training optimizations to determine the most efficient ones.

\begin{table}[H]
\centering
\small
\caption{\small{Optimal parallel strategies for DeepSpeed and Megatron. ``DP'' refers to the data parallel degree, ``TP'' to the tensor parallel degree, ``PP'' to the pipeline parallel degree, and ``SP'' to the sequence parallel (Ulysses-SP) degree. ``AC'' signifies the use of activation checkpointing, while ``bs'' denotes micro-batch size. All DeepSpeed configurations employ ZeRO-3 optimization, whereas Megatron implementations utilize ZeRO-1.}}
\label{tab:hetero_cluster}
\begin{tabular}{lcc}
\hline\toprule
\textbf{Model Size \& GPUs} & \textbf{DeepSpeed} & \textbf{Megatron} \\
\midrule
32B, 16 H800 & DP8SP2AC, bs2 & TP4PP4, bs1 \\
32B, 16 H20 & DP8SP2AC, bs2 & TP4PP4, bs1 \\
32B, 16 H800, 16 H20 & DP16SP2AC, bs2 & DP2TP4PP4, bs2 \\
32B, 16 H800, 24 H20 & DP20SP2AC, bs4 & DP2TP4PP5, bs2 \\
32B, 16 H800, 32 H20 & DP24SP2AC, bs1 & DP4TP4PP3, bs2 \\
70B, 16 H800, 16 H20 & DP16SP2AC, bs1 & TP8PP4, bs1 \\
70B, 16 H800, 24 H20 & DP20SP2AC, bs2 & TP8PP5, bs1 \\
70B, 16 H800, 32 H20 & DP24SP2AC, bs1 & TP8PP6, bs1 \\
\bottomrule\hline
\end{tabular}
\end{table}

Next, as shown in Table~\ref{tab:hetero_cluster_HSPMD}, we sequentially present the optimal strategies for HSPMD across different heterogeneous device setups. These strategies are given by our static planner (as introduced in Appendix~\ref{appendix:plan}). Each cell represents a tensor parallel group, detailing the assigned rank IDs (denoted as ``R'') and the corresponding layer IDs (denoted as ``L''). Yellow cells correspond to H800 GPUs, while white ones represent H20 GPUs. The leftmost column indicates the micro-batch size (denoted as ``bs'') and the number of micro-batches (represented by the preceding coefficient) for each pipeline. Between pipelines, data parallelism is utilized to synchronize gradients.

Notably, across all strategies, we utilize ZeRO-1 to partition FP32 parameters and optimizer states across data parallel replicas. This memory optimization enables the adoption of more memory-intensive parallel strategies (i.e., those with higher data parallel degrees). While homogeneous parallel strategies use standard \texttt{AG} (all-gather) for parameter collection and \texttt{RS} (reduce-scatter) for gradient synchronization, our heterogeneous approach requires specialized operators. Specifically, when tensor parallel degrees differ across data parallel groups, the inserted \texttt{Reshard} operators (responsible for parameter collection and gradient synchronization) are instantiated with \texttt{SplitAG} and \texttt{SplitRS} operators (detailed in \S\ref{subsec:top_comm}). This adaptation facilitates the asymmetric communication required when integrating ZeRO-1 with heterogeneous pipelines.


\begin{table}[H]
\centering
\small
\caption{\small{Optimal parallel strategies for HSPMD on different heterogeneous clusters. R0-15 are H800 GPUs and R16-47 are H20 GPUs.}}
\label{tab:hetero_cluster_HSPMD}
\begin{tabular}{|c|c|c|c|c|}
\hline
\multicolumn{5}{|c|}{32B, 16 H800, 16 H20} \\
\hhline{|*5-}
\multirow{2}{*}{\makecell{32 $\times$ bs1\\(4 Stages)}} &
{R16-19} &
{R20-23} &
\cellcolor{yellow!10}{R0-3} & 
\cellcolor{yellow!10}{R4-7} 
\\
& 
{L0-6} & 
{L7-13} & 
\cellcolor{yellow!10}{L14-36} & 
\cellcolor{yellow!10}{L37-59} 
\\
\hhline{|*5-}
\multirow{2}{*}{\makecell{32 $\times$ bs1\\(4 Stages)}} &
{R24-27} &
{R28-31} &
\cellcolor{yellow!10}{R8-11} & 
\cellcolor{yellow!10}{R12-15} 
\\
& 
{L0-6} & 
{L7-13} & 
\cellcolor{yellow!10}{L14-36} & 
\cellcolor{yellow!10}{L37-59} 
\\ 
\hhline{|*5-}
\end{tabular}


\medskip

\begin{tabular}{|c|c|c|c|c|}
\hline
\multicolumn{5}{|c|}{32B, 16 H800, 24 H20} \\
\hhline{|*5-}
\multirow{4}{*}{\makecell{32 $\times$ bs1\\(5 Stages)}} &
R16-19 &
R20-23 &
R24-27 &
\cellcolor{yellow!10}{R0-3}
\\
& 
L0-5 & 
L6-11 & 
L12-17 & 
\cellcolor{yellow!10}{L18-38}
\\
\hhline{~|*4-}
& 
\cellcolor{yellow!10}{R4-7}
\\
&
\cellcolor{yellow!10}{L39-59} 
\\
\hhline{|*5-}
\multirow{4}{*}{\makecell{32 $\times$ bs1\\(5 Stages)}} &
R28-31 & 
R32-35 & 
R36-39 &
\cellcolor{yellow!10}{R8-11}
\\
& 
L0-5 & 
L6-11 & 
L12-17 & 
\cellcolor{yellow!10}{L18-38}
\\
\hhline{~|*4-}
& 
\cellcolor{yellow!10}{R12-15}
\\
&
\cellcolor{yellow!10}{L39-59} 
\\
\hhline{|*2-}
\end{tabular}


\medskip

\begin{tabular}{|c|c|c|c|}
\hline
\multicolumn{4}{|c|}{32B, 16 H800, 32 H20} \\
\hhline{|*4-}
\multirow{2}{*}{\makecell{16 $\times$ bs1\\(3 Stages)}} &
{R16-19} &
{R20-23} &
\cellcolor{yellow!10}{R0-3} 
\\
& 
{L0-10} & 
{L11-21} & 
\cellcolor{yellow!10}{L22-59} 
\\
\hhline{|*4-}
\multirow{2}{*}{\makecell{16 $\times$ bs1\\(3 Stages)}} &
{R24-27} &
{R28-31} &
\cellcolor{yellow!10}{R4-7} 
\\
& 
{L0-10} & 
{L11-21} & 
\cellcolor{yellow!10}{L22-59} 
\\
\hhline{|*4-}
\multirow{2}{*}{\makecell{16 $\times$ bs1\\(3 Stages)}} &
{R32-35} &
{R36-39} &
\cellcolor{yellow!10}{R8-11} 
\\
& 
{L0-10} & 
{L11-21} & 
\cellcolor{yellow!10}{L22-59} 
\\
\hhline{|*4-}
\multirow{2}{*}{\makecell{16 $\times$ bs1\\(3 Stages)}} &
{R40-43} &
{R44-47} &
\cellcolor{yellow!10}{R12-15} 
\\
& 
{L0-10} & 
{L11-21} & 
\cellcolor{yellow!10}{L22-59} 
\\
\hhline{|*4-}
\end{tabular}


\medskip

\begin{tabular}{|c|c|c|c|c|}
\hline
\multicolumn{5}{|c|}{70B, 16 H800, 16 H20} \\
\hhline{|*5-}
\multirow{2}{*}{\makecell{64 $\times$ bs1\\(4 Stages)}} &
{R16-23} &
{R24-31} &
\cellcolor{yellow!10}{R0-7} & 
\cellcolor{yellow!10}{R8-15} 
\\
& 
{L0-10} & 
{L11-21} & 
\cellcolor{yellow!10}{L22-50} & 
\cellcolor{yellow!10}{L51-79} 
\\
\hhline{|*5-}
\end{tabular}


\medskip

\begin{tabular}{|c|c|c|c|c|}
\hline
\multicolumn{5}{|c|}{70B, 16 H800, 24 H20} \\
\hhline{|*5-}
\multirow{4}{*}{\makecell{64 $\times$ bs1\\(5 Stages)}} &
R16-23 &
R24-31 &
R32-39 &
\cellcolor{yellow!10}{R0-7}
\\
& 
L0-9 & 
L10-19 & 
L20-29 & 
\cellcolor{yellow!10}{L30-54}
\\
\hhline{~|*4-}
& 
\cellcolor{yellow!10}{R8-15}
\\
&
\cellcolor{yellow!10}{L55-79} 
\\
\hhline{|*2-}
\end{tabular}


\medskip

\begin{tabular}{|c|c|c|c|}
\hline
\multicolumn{4}{|c|}{70B, 16 H800, 32 H20} \\
\hhline{|*4-}
\multirow{2}{*}{\makecell{32 $\times$ bs1\\(3 Stages)}} &
{R16-23} &
{R24-31} &
\cellcolor{yellow!10}{R0-7} 
\\
& 
{L0-16} & 
{L17-33} & 
\cellcolor{yellow!10}{L34-79} 
\\
\hhline{|*4-}
\multirow{2}{*}{\makecell{32 $\times$ bs1\\(3 Stages)}} &
{R32-39} &
{R40-47} &
\cellcolor{yellow!10}{R8-15} 
\\
& 
{L0-16} & 
{L17-33} & 
\cellcolor{yellow!10}{L34-79} 
\\
\hhline{|*4-}
\end{tabular}
\end{table}

\subsection{Scenario (b): Unstable Devices}
\label{appendix:elastic}

Table~\ref{tab:elastic} presents the optimal strategies of DeepSpeed and Megatron after each reconfiguration in the unstable devices scenario. Notably, GPU failures occurred during the reconfigurations from C1 to C2 and from C5 to C6. However, due to the system's inability to support non-uniform sharding, they failed to utilize the remaining seven functional GPUs in the machine. Consequently, the entire machine was treated as failed, which is why the strategy for C2 matches C3, and C6's strategy aligns with C7's.

\begin{table}[H]
\centering
\small
\caption{\small{Optimal parallel strategies for DeepSpeed and Megatron. ``DP'' refers to the data parallel degree, ``TP'' to the tensor parallel degree, ``PP'' to the pipeline parallel degree, and ``SP'' to the sequence parallel (Ulysses-SP) degree. ``AC'' signifies the use of activation checkpointing, while ``bs'' denotes micro-batch size. All DeepSpeed configurations employ ZeRO-3 optimization, whereas Megatron implementations utilize ZeRO-1.}}
\label{tab:elastic}
\begin{tabular}{lcc}
\hline\toprule
\textbf{Configuration} & \textbf{DeepSpeed} & \textbf{Megatron} \\
\midrule
C1: 32 H20 & DP16SP2AC, bs2 & DP2TP4PP4, bs2 \\
C2: 31 H20 & DP12SP2AC, bs2 & TP4PP6, bs1 \\
C3: 24 H20 & DP12SP2AC, bs2 & TP4PP6, bs1 \\
C4: 16 H800, 32 H20 &  DP24SP2AC, bs1 & DP4TP4PP3, bs2 \\
C5: 16 H800, 24 H20 & DP20SP2AC, bs2 & TP8PP5, bs1 \\
C6: 15 H800, 24 H20 & DP16SP2AC, bs2 & DP2TP4PP4, bs2 \\
C7: 8 H800, 24 H20 & DP16SP2AC, bs2 & DP2TP4PP4, bs2 \\
\bottomrule\hline
\end{tabular}
\end{table}

Upon changes in GPU availability, we invoke our online planner (Appendix~\ref{appendix:plan}) to determine the optimal parallel strategy.
Table~\ref{tab:elastic_homo_HSPMD} reports the strategies adopted by HSPMD on homogeneous clusters (C1–C3). In our setup, we deploy two pipelines and disable ZeRO-1 to ensure fault isolation. Failures in one pipeline do not cause permanent loss of model weights, eliminating the need for restarting. This setting is consistent with Oobleck, which also disables ZeRO-1 and achieves fault tolerance through data-parallel replication, making our comparison fair. Notably, the same design is followed by many elastic training systems (e.g., Recycle~\cite{ReCycle}).


\begin{table}[H]
\centering
\small
\caption{\small{Parallel strategies for HSPMD during elastic training on homogeneous clusters.}}
\label{tab:elastic_homo_HSPMD}
\begin{tabular}{|c|c|c|c|c|}
\hline
\multicolumn{5}{|c|}{C1: 32 H20} \\
\hhline{|*5-}
\multirow{2}{*}{\makecell{16 $\times$ bs2\\(4 Stages)}} &
{R0-3} &
{R4-7} &
{R8-11} & 
{R12-15} 
\\
& 
{L0-14} & 
{L15-29} & 
{L30-44} & 
{L45-59} 
\\
\hhline{|*5-}
\multirow{2}{*}{\makecell{16 $\times$ bs2\\(4 Stages)}} &
{R16-19} &
{R20-23} &
{R24-27} & 
{R28-31} 
\\
& 
{L0-14} & 
{L15-29} & 
{L30-44} & 
{L45-59}  
\\ 
\hhline{|*5-}
\end{tabular}


\medskip

\begin{tabular}{|c|c|c|c|c|}
\hline
\multicolumn{5}{|c|}{C2: 31 H20} \\
\hhline{|*5-}
\multirow{2}{*}{\makecell{33 $\times$ bs1\\(4 Stages)}} &
{R0-3} &
{R4-7} &
{R8-11} & 
{R12-15} 
\\
& 
{L0-14} & 
{L15-29} & 
{L30-44} & 
{L45-59} 
\\
\hhline{|*5-}
\multirow{4}{*}{\makecell{31 $\times$ bs1\\(5 Stages)}} &
{R16-19} &
{R20-23} &
{R24-27} & 
{R28-29} 
\\
& 
L0-15 & 
L16-31 & 
L32-47 & 
L48-55
\\
\hhline{~|*4-}
& 
R30
\\
&
L56-59 
\\
\hhline{|*2-}
\end{tabular}


\medskip

\begin{tabular}{|c|c|c|c|}
\hline
\multicolumn{4}{|c|}{C3: 24 H20} \\
\hhline{|*4-}
\multirow{2}{*}{\makecell{32 $\times$ bs1\\(3 Stages)}} &
R0-3 &
R4-7 &
R8-11
\\
& 
{L0-19} & 
{L20-39} & 
{L40-59} 
\\
\hhline{|*4-}
\multirow{2}{*}{\makecell{32 $\times$ bs1\\(3 Stages)}} &
{R12-15} &
{R16-19} &
{R20-23} 
\\
& 
{L0-19} & 
{L20-39} & 
{L40-59} 
\\
\hhline{|*4-}
\end{tabular}
\end{table}

In Table~\ref{tab:elastic_hetero_HSPMD}, we detail the strategies (C4–C7) adopted by HSPMD on heterogeneous clusters, with yellow and white cells representing H800 and H20 GPUs, respectively. Following the homogeneous cluster approach, we disable ZeRO-1 and configure two pipelines to enable fault tolerance.

However, disabling ZeRO-1 increases memory overhead, precluding the use of the optimal parallel strategy. For instance, the strategy in C4: 16 H800, 32 H20 (with a data parallel degree of 2) differs from that in Table~\ref{tab:hetero_cluster_HSPMD} for standard heterogeneous clusters (32B, 16 H800, 32 H20, with a data parallel degree of 4). Consequently, training performance degrades by approximately 15\%, increasing the per-step training time from 6.05s to 6.91s (shown in Figure~\ref{fig:mfu} and Figure~\ref{fig:elastic}).

Despite this performance degradation, HSPMD’s heterogeneous sharding effectively balances workloads, still maintaining superior performance compared to the homogeneous parallel strategies used in DeepSpeed and Megatron (shown in Figure~\ref{fig:elastic}).

\begin{table}[H]
\centering
\small
\caption{\small{Parallel strategies for HSPMD during elastic training on heterogeneous clusters. R0-15 are H800 GPUs and R16-47 are H20 GPUs.}}
\label{tab:elastic_hetero_HSPMD}
\begin{tabular}{|c|c|c|c|c|}
\hline
\multicolumn{5}{|c|}{C4: 16 H800, 32 H20} \\
\hhline{|*5-}
\multirow{4}{*}{\makecell{32 $\times$ bs1\\(6 Stages)}} &
R16-19 &
R20-23 &
R24-27 &
R28-31
\\
& 
L0-4 & 
L5-10 & 
L11-16 & 
{L17-22}
\\
\hhline{~|*4-}
& 
\cellcolor{yellow!10}{R0-3} &
\cellcolor{yellow!10}{R4-7}
\\
&
\cellcolor{yellow!10}{L23-40} &
\cellcolor{yellow!10}{L41-59}
\\
\hhline{|*5-}
\multirow{4}{*}{\makecell{32 $\times$ bs1\\(6 Stages)}} &
R32-35 & 
R36-39 & 
R40-43 &
R44-47
\\
& 
L0-4 & 
L5-10 & 
L11-16 & 
{L17-22}
\\
\hhline{~|*4-}
& 
\cellcolor{yellow!10}{R8-11} &
\cellcolor{yellow!10}{R12-15}
\\
&
\cellcolor{yellow!10}{L23-40} &
\cellcolor{yellow!10}{L41-59}
\\
\hhline{|*3-}
\end{tabular}


\medskip

\begin{tabular}{|c|c|c|c|c|}
\hline
\multicolumn{5}{|c|}{C5: 16 H800, 24 H20} \\
\hhline{|*5-}
\multirow{4}{*}{\makecell{32 $\times$ bs1\\(5 Stages)}} &
R16-19 &
R20-23 &
R24-27 &
\cellcolor{yellow!10}{R0-3}
\\
& 
L0-5 & 
L6-11 & 
L12-17 & 
\cellcolor{yellow!10}{L18-38}
\\
\hhline{~|*4-}
& 
\cellcolor{yellow!10}{R4-7}
\\
&
\cellcolor{yellow!10}{L39-59} 
\\
\hhline{|*5-}
\multirow{4}{*}{\makecell{32 $\times$ bs1\\(5 Stages)}} &
R28-31 & 
R32-35 & 
R36-39 &
\cellcolor{yellow!10}{R8-11}
\\
& 
L0-5 & 
L6-11 & 
L12-17 & 
\cellcolor{yellow!10}{L18-38}
\\
\hhline{~|*4-}
& 
\cellcolor{yellow!10}{R12-15}
\\
&
\cellcolor{yellow!10}{L39-59} 
\\
\hhline{|*2-}
\end{tabular}


\medskip

\begin{tabular}{|c|c|c|c|c|}
\hline
\multicolumn{5}{|c|}{C6: 15 H800, 24 H20} \\
\hhline{|*5-}
\multirow{4}{*}{\makecell{33 $\times$ bs1\\(5 Stages)}} &
R16-19 &
R20-23 &
R24-27 &
\cellcolor{yellow!10}{R0-3}
\\
& 
L0-5 & 
L6-11 & 
L12-17 & 
\cellcolor{yellow!10}{L18-38}
\\
\hhline{~|*4-}
& 
\cellcolor{yellow!10}{R4-7}
\\
&
\cellcolor{yellow!10}{L39-59} 
\\
\hhline{|*5-}
\multirow{4}{*}{\makecell{31 $\times$ bs1\\(6 Stages)}} &
R28-31 & 
R32-35 & 
R36-39 &
\cellcolor{yellow!10}{R8-11}
\\
& 
L0-5 & 
L6-11 & 
L12-17 & 
\cellcolor{yellow!10}{L18-39}
\\
\hhline{~|*4-}
& 
\cellcolor{yellow!10}{R12-13} &
\cellcolor{yellow!10}{R14}
\\
&
\cellcolor{yellow!10}{L40-52} &
\cellcolor{yellow!10}{L53-59}
\\
\hhline{|*3-}
\end{tabular}


\medskip

\begin{tabular}{|c|c|c|c|c|}
\hline
\multicolumn{5}{|c|}{C7: 8 H800, 24 H20} \\
\hhline{|*5-}
\multirow{2}{*}{\makecell{32 $\times$ bs1\\(4 Stages)}} &
{R16-19} &
{R20-23} &
{R24-27} & 
\cellcolor{yellow!10}{R0-3} 
\\
& 
{L0-8} & 
{L9-18} & 
{L19-28} & 
\cellcolor{yellow!10}{L29-59} 
\\
\hhline{|*5-}
\multirow{2}{*}{\makecell{32 $\times$ bs1\\(4 Stages)}} &
{R28-31} &
{R32-35} &
{R36-39} & 
\cellcolor{yellow!10}{R4-7} 
\\
& 
{L0-8} & 
{L9-18} & 
{L19-28} & 
\cellcolor{yellow!10}{L29-59} 
\\ 
\hhline{|*5-}
\end{tabular}

\end{table}

\subsection{Scenario (c): Mixed-length Data}
\label{appendix:mixed}

Table~\ref{tab:mix} presents the optimal parallel training strategies for the mixed-length data scenario using DeepSpeed and Megatron, with context lengths of 32K and 16K (using the 32B model on 32 H20 GPUs). Due to the substantially longer context lengths, these frameworks must adopt larger sequence/context or tensor parallelism. However, in practice, only a small fraction of sequences per training step actually reach the maximum context length. As a result, most shorter sequences are processed inefficiently, missing opportunities for more optimized training strategies.

\begin{table}[H]
\centering
\small
\caption{\small{Optimal parallel strategies for DeepSpeed and Megatron. ``DP'' refers to the data parallel degree, ``TP'' to the tensor parallel degree, ``PP'' to the pipeline parallel degree, ``CP'' to the context parallel degree, and ``SP'' to the sequence parallel (Ulysses-SP) degree. ``AC'' signifies the use of activation checkpointing, while ``bs'' denotes micro-batch size. All DeepSpeed configurations employ ZeRO-3 optimization, whereas Megatron implementations utilize ZeRO-1.}}
\label{tab:mix}
\begin{tabular}{lcc}
\hline\toprule
\textbf{Context Length} & \textbf{DeepSpeed} & \textbf{Megatron} \\
\midrule
32K & DP4SP8AC, bs1 & DP2TP8CP2, bs1 \\
16K & DP8SP4AC, bs1 & TP8PP4, bs1 \\
\bottomrule\hline
\end{tabular}
\end{table}

Table~\ref{tab:mix_hotspa} presents the parallel strategies adopted by HotSPa across different sequence length intervals. By partitioning sequences into buckets based on length and processing them sequentially within a training step (with gradient accumulation before weight updates), this approach maintains mathematical equivalence to standard training while enabling length-adaptive parallel strategies.

Although this method incurs frequent switching overhead within the step, such costs are substantially outweighed by the benefits of parallel strategies optimization, ultimately resulting in performance improvements.

\begin{table}[H]
\centering
\small
\caption{\small{Optimal parallel strategies for HotSPa. ``DP'' refers to the data parallel degree, ``TP'' to the tensor parallel degree, and ``PP'' to the pipeline parallel degree. ``bs'' denotes micro-batch size. All configurations employ ZeRO-1 optimization.}}
\label{tab:mix_hotspa}
\begin{tabular}{lc}
\hline\toprule
\textbf{Context Length} & \textbf{Interval-based Parallel Strategies} \\
\midrule
\multirow{3}{*}{\makecell{32K}} & 16K-32K: DP2TP16, bs1 \\
& 4K-16K: DP2TP8PP2, bs1 \\
& 0-4K: DP4TP4PP2, bs1 \\
\midrule
\multirow{2}{*}{\makecell{16K}} &
4K-16K: DP2TP8PP2, bs1 \\
& 0-4K: DP4TP4PP2, bs1 \\
\bottomrule\hline
\end{tabular}
\end{table}

Tables~\ref{tab:mix_HSPMD_32K} and~\ref{tab:mix_HSPMD_16K} further present the heterogeneous strategies adopted by HSPMD for context lengths of 32K and 16K, respectively. In Table~\ref{tab:mix_HSPMD_32K}, Strategy A and Strategy B correspond to those illustrated in Figure~\ref{fig:seq_and_time}.

For HSPMD, steps with varying maximum sequence lengths are processed using either Strategy A or B, while sequences of different lengths within a step are distributed across multiple pipelines to balance the load based on the \textit{cost model} we designed. Once all data within a step has been processed across the pipelines, gradients are synchronized to perform the step update. This approach further improves performance compared to HotSPa by eliminating frequent strategy switching within a step and more effectively addressing workload imbalance through the use of heterogeneous strategies, rather than homogeneous ones.

\begin{table}[H]
\centering
\small
\caption{\small{Heterogeneous parallel strategies for HSPMD with 32K context length, where the strategy is dynamically selected based on the maximum sequence length (MaxSeqLen) at each processing step.}}
\label{tab:mix_HSPMD_32K}
\begin{tabular}{|c|c|}
\hline
\multicolumn{2}{|c|}{Strategy A: MaxSeqLen $\in$ (16K, 32K]} \\
\hhline{|*2-}
\multirow{2}{*}{\makecell{Long Sequence Pipeline-1\\(1 Stages)}} &
{R0-15}
\\
& 
{L0-59} 
\\
\hhline{|*2-}
\multirow{2}{*}{\makecell{Short Sequence Pipeline-1\\(1 Stages)}} &
R16-19
\\
& 
{L0-59} 
\\ 
\hhline{|*2-}
\multirow{2}{*}{\makecell{...\\(1 Stages)}} &
...
\\
& 
{L0-59} 
\\ 
\hhline{|*2-}
\multirow{2}{*}{\makecell{Short Sequence Pipeline-4\\(1 Stages)}} &
R28-31
\\
& 
{L0-59} 
\\ 
\hhline{|*2-}
\end{tabular}

\medskip

\begin{tabular}{|c|c|c|}
\hline
\multicolumn{3}{|c|}{Strategy B: MaxSeqLen $\in$ (0K, 16K]} \\
\hhline{|*3-}
\multirow{2}{*}{\makecell{Long Sequence Pipeline-1\\(1 Stages)}} &
{R0-7}
\\
& 
{L0-59} 
\\
\hhline{|*3-}
\multirow{2}{*}{\makecell{Short Sequence Pipeline-1\\(2 Stages)}} &
R8-11 &
R12-15
\\
& 
{L0-29} &
{L30-59}
\\ 
\hhline{|*3-}
\multirow{2}{*}{\makecell{...\\(2 Stages)}} &
... &
...
\\
& 
{L0-29} &
{L30-59}
\\ 
\hhline{|*3-}
\multirow{2}{*}{\makecell{Short Sequence Pipeline-3\\(2 Stages)}} &
R24-27 &
R28-31
\\
& 
{L0-29} &
{L30-59}
\\ 
\hhline{|*3-}
\end{tabular}

\end{table}

\begin{table}[H]
\centering
\small
\caption{\small{Heterogeneous parallel strategies for HSPMD with 16K context length, where the strategy is dynamically selected based on the maximum sequence length (MaxSeqLen) at each processing step.}}
\label{tab:mix_HSPMD_16K}
\begin{tabular}{|c|c|c|}
\hline
\multicolumn{3}{|c|}{Strategy A: MaxSeqLen $\in$ (4K, 16K]} \\
\hhline{|*3-}
\multirow{2}{*}{\makecell{Long Sequence Pipeline-1\\(1 Stages)}} &
{R0-7}
\\
& 
{L0-59} 
\\
\hhline{|*3-}
\multirow{2}{*}{\makecell{Short Sequence Pipeline-1\\(2 Stages)}} &
R8-11 &
R12-15
\\
& 
{L0-29} &
{L30-59}
\\ 
\hhline{|*3-}
\multirow{2}{*}{\makecell{...\\(2 Stages)}} &
... &
...
\\
& 
{L0-29} &
{L30-59}
\\ 
\hhline{|*3-}
\multirow{2}{*}{\makecell{Short Sequence Pipeline-3\\(2 Stages)}} &
R24-27 &
R28-31
\\
& 
{L0-29} &
{L30-59}
\\ 
\hhline{|*3-}
\end{tabular}

\medskip

\begin{tabular}{|c|}
\hline
\multicolumn{1}{|c|}{Strategy B: MaxSeqLen $\in$ (0K, 4K]} \\ \hhline{|*1-}
DP4TP4PP2 \\
\hhline{|*1-}
\end{tabular}

\end{table}


\end{document}